\documentclass[journal]{IEEEtran}
\usepackage{amsmath,amssymb,amsfonts}
\usepackage{algorithmic}
\usepackage{graphicx}
\usepackage{algorithm,algorithmic}
\usepackage{hyperref}
\hypersetup{hidelinks=true}
\usepackage{textcomp}
\usepackage{multirow}
\usepackage{tikz}
\usepackage{svg}
\usepackage{siunitx,booktabs}
\usepackage{soul}
\usepackage{graphicx}
\usepackage{caption}
\usetikzlibrary{trees,positioning,shapes,shadows,arrows.meta}
\usepackage[edges]{forest}
\usepackage[
sorting=nty,
backend=bibtex,
style=ieee,
sorting=none,
defernumbers=false]{biblatex}
\def\BibTeX{{\rm B\kern-.05em{\sc i\kern-.025em b}\kern-.08em
    T\kern-.1667em\lower.7ex\hbox{E}\kern-.125emX}}

\usepackage{titlesec}

\begin{document}
\title{Deep Learning for Ophthalmology: \\ The State-of-the-Art and Future Trends}


\author{Duy M. H. Nguyen$^\dagger$, Hasan Md Tusfiqur Alam$^\dagger$,  Tai Nguyen, Devansh Srivastav \\ Hans-Juergen Profitlich, Ngan Le, and Daniel Sonntag
\thanks{$^\dagger$ Equal contribution}
\thanks{Duy M. H. Nguyen is with the German Research Center for Artificial Intelligence (DFKI), International Max Planck Research School for Intelligent Systems (IMPRS-IS), and Department of Computer Science, University of Stuttgart, Germany.}
\thanks{Hasan Md Tusfiqur Alam is with the German Research Center for Artificial Intelligence (DFKI).}
\thanks{Tai Nguyen is with the German Research Center for Artificial Intelligence (DFKI).}
\thanks{Devansh Srivastav is with the German Research Center for Artificial Intelligence (DFKI).}
\thanks{Hans-Juergen Profitlich is with the German Research Center for Artificial Intelligence (DFKI).}
\thanks{Ngan Le is with the Department of Computer Science and Computer Engineering, University
of Arkansas, USA.}
\thanks{Daniel Sonntag is with the German Research Center for Artificial Intelligence (DFKI) and the Department of Applied Artificial Intelligence, Oldenburg University, Germany.}
}

\maketitle
\begin{abstract}
The emergence of artificial intelligence (AI), particularly deep learning (DL), has marked a new era in the realm of ophthalmology, offering transformative potential for the diagnosis and treatment of posterior segment eye diseases. 
This review explores the cutting-edge applications of DL across a range of ocular conditions, including diabetic retinopathy, glaucoma, age-related macular degeneration, and retinal vessel segmentation. We provide a comprehensive overview of foundational ML techniques and advanced DL architectures, such as CNNs, attention mechanisms, and transformer-based models, highlighting the evolving role of AI in enhancing diagnostic accuracy, optimizing treatment strategies, and improving overall patient care.
Additionally, we present key challenges in integrating AI solutions into clinical practice, including ensuring data diversity, improving algorithm transparency, and effectively leveraging multimodal data. This review emphasizes AI’s potential to improve disease diagnosis and enhance patient care while stressing the importance of collaborative efforts to overcome these barriers and fully harness AI’s impact in advancing eye care.
\end{abstract}




\section{\textbf{Introduction}}
\label{sec:introduction}

\begin{figure}[!hbt]
    \centering
    \includegraphics[width=0.4\textwidth]{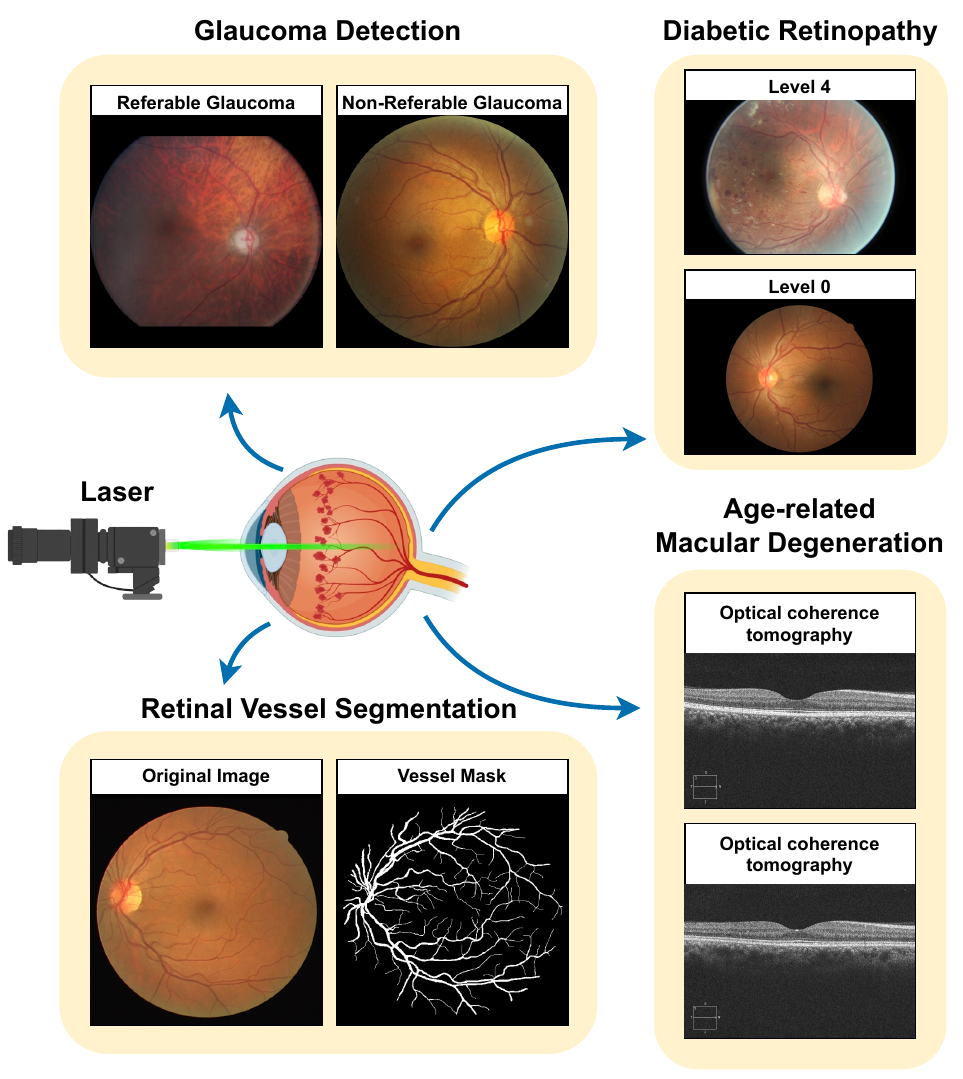}
    \caption{Overall schematic diagram describing four main problems in common ophthalmic imaging modalities presented in our survey.}  
    \label{fig:retinal_problem}
\end{figure}
\IEEEPARstart{A}{rtificial} intelligence encompasses algorithms and tools that replicate human intelligence digitally, drawing from disciplines such as logic, computer science, and psychology \cite{winston1992artificial,ertel2024introduction}. Its applications range from voice recognition to intelligent robotics \cite{russell2016artificial}, with significant potential in healthcare \cite{jiang2017artificial,secinaro2021role}. AI utilizes machine learning (ML) \cite{jordan2015machine} and deep learning (DL) \cite{lecun2015deep} techniques to accelerate automation. Recent advances in DL \cite{lecun2015deep} include models like convolutional neural networks (CNNs), recurrent neural networks (RNNs), autoencoders (AEs), and transformers. CNNs, a cornerstone in supervised DL \cite{lecun1998gradient,bouvrie2006notes,krizhevsky2012imagenet}, are primarily used for image classification, object detection, and segmentation, comprising convolutional, pooling, and fully connected layers \cite{li2021survey}. RNNs \cite{rumelhart1986learning,schmidt2019recurrent}, crucial for sequential data like speech and text, use cyclic hidden units for recurrent computations, with variants such as long-short term memory (LSTM) \cite{graves2012long} and gated recurrent units (GRUs) \cite{cho2014properties}. AEs focus on efficient data coding \cite{bengio2013representation}, mapping input data to itself for feature reduction and network initialization. Transformers, based on self-attention mechanisms, capture long-term dependencies in sequences \cite{vaswani2017attention}, excelling in Natural Language Processing (NLP) \cite{devlin2018bert} and vision tasks like the vision transformers (ViTs) \cite{dosovitskiy2020image}. However, the attention mechanism in transformers requires substantial GPU memory.

The integration of advanced AI, particularly DL, marks a transformative shift in the field of medical diagnostics and treatment. By enabling unprecedented levels of precision, scalability, and efficiency, DL technologies are poised to reshape traditional healthcare paradigms, ultimately improving patient outcomes and operational workflows. Ophthalmology, with its heavy reliance on visual data for disease detection and monitoring, stands out as a key beneficiary of these advancements. The ability of DL algorithms to process and interpret complex medical imaging data, such as fundus photographs, optical coherence tomography (OCT), and slit-lamp images, offers tremendous potential for enhancing the diagnosis and management of ocular diseases. Research has already demonstrated DL’s capability to surpass human-level performance in tasks such as detecting diabetic retinopathy, age-related macular degeneration, and glaucoma \cite{ting2017development, de2018clinically}. This efficiency stems from its ability to rapidly and accurately analyze high-dimensional image data while identifying subtle patterns often imperceptible to the human eye.

The strengths of DL are not only as a diagnostic tool but also as a means of improving treatment planning, including the personalization of patient care. Currently, developed prediction models can forecast disease progression and response to treatment, thereby assisting ophthalmologists in tailoring treatment plans to individual patient predictions \cite{bogunovic2017machine}. For instance, DL algorithms have demonstrated high efficacy in analyzing complex datasets from imaging techniques such as fundus photography and OCT, improving diagnostic accuracy for conditions like diabetic retinopathy, glaucoma, and age-related macular degeneration \cite{ting2019artificial}. Moreover, AI-driven predictive models are being developed to anticipate the progression of diseases such as glaucoma by analyzing longitudinal patient data, which enables early intervention and personalized treatment planning \cite{li2023medical}. This capability not only enhances patient outcomes but also optimizes healthcare resources by focusing on preventive care. 

However, applying DL in ophthalmology presents several significant challenges, particularly in areas such as human interaction, XAI, multi-modal approaches, and data privacy. Clinicians and patients may be hesitant to trust AI systems, and seamlessly integrating these tools into existing workflows without causing disruption is complex \cite{ting2019artificial}. Moreover, the ``black box'' nature of DL models raises concerns about transparency and the ability to explain AI decisions, which are critical for regulatory compliance and ethical standards \cite{singh2021evaluation,castelvecchi2016can}. Multi-modal approaches, which aim to combine data from various sources, face technical difficulties in data integration, risk overfitting, and require substantial computational resources \cite{mossdeep,frasca2024explainable}. Compounding these issues is the paramount need to ensure data privacy, as AI systems often handle sensitive patient information. To maintain trust and safeguard this data, it is essential to implement robust security measures and establish comprehensive regulatory frameworks that prioritize patient confidentiality and data protection \cite{naresh2023privacy,beaulieu2018privacy,tom2020protecting}.


Our objective is to present a comprehensive review of cutting-edge AI techniques in ophthalmology, focusing on key applications such as diabetic retinopathy, glaucoma, age-related macular degeneration (AMD), and vessel segmentation. The review delves into methodologies, traditional approaches, state-of-the-art neural network architectures, performance metrics, and supporting datasets. The paper is organized as follows: (i) an overview of ophthalmology, problem formulation, and related literature in Section \ref{sec:background}; (ii) an in-depth analysis of techniques across various ophthalmology applications, including methodologies and performance evaluations with existing datasets in Section \ref{sec:applications}; (iii) a discussion on fundamental challenges and future opportunities for AI in ophthalmology in Section \ref{sec:future}; and (iv) concluding remarks in Section \ref{sec:conclusion}.

\begin{figure}[!ht]
    \centering
    \includegraphics[width=0.44\textwidth]{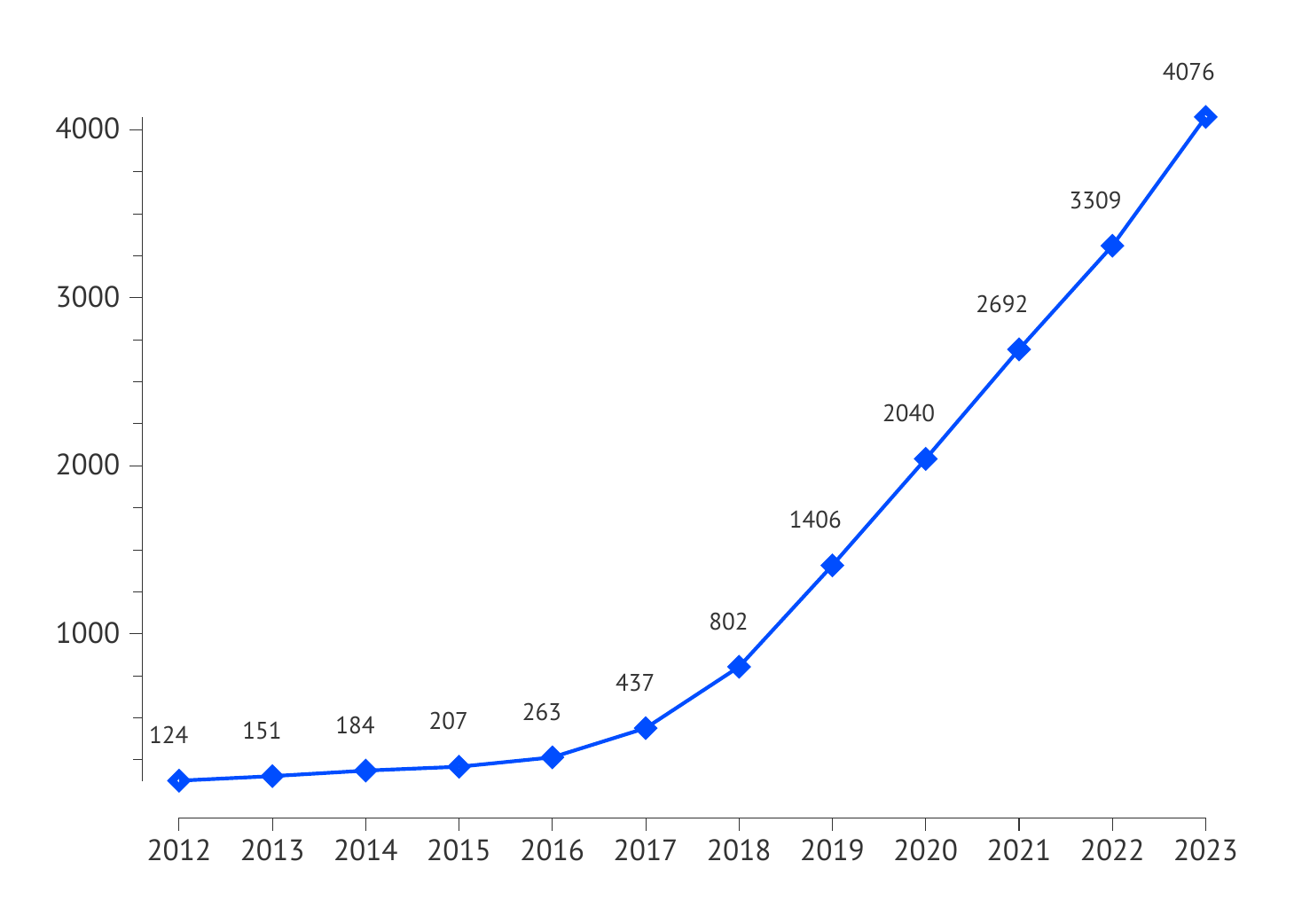}
    \caption{Rapid growth in publications leveraging machine learning and deep learning for ophthalmology from 2012 to 2023.}
    \label{fig:pubstat}
\end{figure}

Compared to previous surveys \cite{ting2019artificial,ting2019deep,tong2020application,abbas2022machine,linde2024comparative,litao2021,9783206, ILESANMI2023100261,10.1007/s11042-023-15110-9,PAVITHRA2023157}, our work offers a significantly broader scope by encompassing a wider range of DL applications in ophthalmology. While earlier studies primarily focus on specific areas or methodologies, we aim to provide a holistic overview that captures the diversity and depth of recent advancements in this field. Notably, we delve into emerging topics such as transformer models, which have revolutionized natural language processing and are now being increasingly applied in medical imaging, as well as multimodal learning approaches that integrate data from various sources to enhance predictive performance. Furthermore, we go beyond textual analysis by presenting a comprehensive summary figure (Figure \ref{fig:pubstat}) that visually depicts publication trends over the past twelve years. This figure highlights the rapid growth and evolving focus of research in DL for ophthalmology, offering valuable insights into the field’s development.

\section{\textbf{Backgrounds}}
\label{sec:background}

\textbf{Motivation}: 

With the rising prevalence of posterior-segment eye diseases (PSED) such as glaucoma, macular degeneration, and diabetic retinopathy \cite{Flaxman2017}, there is an urgent need for comprehensive and up-to-date insights into how artificial intelligence (AI) can address these challenges. These diseases are major contributors to global vision impairment and blindness, underscoring the importance of early detection and effective management as public health priorities. Aligning with the World Health Organization’s (WHO) 2030 targets to reduce the burden of vision loss \cite{who2022}, it is essential to explore and document the latest advancements in AI applications for PSED. The rapid evolution of AI techniques, including innovations in explainable AI (XAI), multimodal learning, and automatic retinal vessel segmentation, necessitates a fresh synthesis of current progress. An up-to-date survey, therefore not only highlights the transformative potential of these technologies but also identifies gaps, fosters collaboration, and guides future research efforts to improve diagnostic accuracy, treatment outcomes, and accessibility in ophthalmology

\vspace{0.1in}
\textbf{Methodology}:

Thanks to public scientific databases such as PubMed\footnote{\url{https://pubmed.ncbi.nlm.nih.gov}}, IEEE\footnote{\url{https://ieeexplore.ieee.org}}, ScienceDirect\footnote{\url{https://www.sciencedirect.com}}, Nature\footnote{\url{https://www.nature.com}}, Springer\footnote{\url{https://link.springer.com}}, and others, we can get access to relevant studies. These platforms offer public search engines that allow filtering by keywords and time range. In Figure \ref{fig:pubstat}, we present a statistic showing the number of publications changing per year from 2012 to 2023, indicating a prevalence of research studies on the application of ML- or DL-based methods in the diagnosis of ophthalmic diseases. In this study, we focus on the successful application of methods from January 2019 to December 2023, concentrating on publications from top-tier venues and prestigious publishers in the fields of computer science and medicine. The selection criteria include papers related to keywords such as \texttt{ retinopathy, diabetes, diabetic retinopathy, glaucoma, Age-related Macular Degeneration (AMD), diabetic macular edema, color fundus photography, OCT, diabetic retinopathy diagnosis, eye-related disorders, and retinal disease}. 
\vspace{-0.1in}
\subsection{\textbf{Problem Formulation and Taxonomy}}

Ophthalmic imaging modalities face several challenges in diagnosing and managing eye diseases effectively. This section presents a comprehensive overview of four main problems: (i) diabetic retinopathy, (ii) glaucoma, (iii) AMD, and (iv) retinal vessel segmentation, as shown in Figure \ref{fig:retinal_problem}. Each subsection explores the severity, prevalence, impact, and evolving diagnostic and therapeutic approaches for these conditions.

\vspace{0.05in}
\subsubsection{\textbf{Diabetic Retinopathy}}
Over the past two decades, the global incidence of diabetes has surged to three times its previous levels, posing a growing public health challenge. Persistently high blood sugar levels associated with diabetes cause significant microvascular damage, leading to complications such as diabetic retinopathy (DR). Affecting approximately 34.6\% of diabetics, DR is a leading cause of vision impairment and blindness among adults aged 20 to 74 \cite{burton2021lancet}. This condition arises from damage to the retinal blood vessels, which can result in swelling, leakage, or abnormal vessel growth. Despite its prevalence, DR remains a preventable cause of blindness, with current treatments such as timely laser therapy and intraocular injections showing promising outcomes \cite{CHEUNG2010124}. Screening thus is crucial, and advances in scanning confocal ophthalmology, teleophthalmology, and AI are improving strategies, cost-effectiveness, and broader roles beyond preventing sight-threatening disease \cite{VUJOSEVIC2020337}.

\vspace{0.05in}
\subsubsection{\textbf{Glaucoma}}
The primary cause of irreversible blindness on a global scale, manifests as a varied set of diseases characterized by optic nerve head cupping and visual-field damage, with a prevalence of approximately 3.5\% in individuals aged 40 or above \cite{Kingf3518}. Early detection through ophthalmological examination is vital, and risk factors differ among various glaucoma types, emphasizing the significance of tailored diagnostic and treatment approaches \cite{JONAS20172183}. The worldwide prevalence of glaucoma is anticipated to notably increase to 111.8 million by 2040, disproportionately affecting populations in Asia and Africa \cite{tham2014global}. These projections highlight the essential requirement for strategic planning in glaucoma screening, treatment, and public health initiatives \cite{THAM20142081}.

\vspace{0.05in}
\subsubsection{\textbf{Age-related Macular Degeneration}}
AMD progresses through various stages, beginning with early signs such as medium-sized drusen and retinal pigmentary changes and advancing to late-stage forms, including neovascular (wet) and atrophic (dry) AMD. The pathogenesis of this complex disease is influenced by dysregulation in multiple biological pathways, including those related to complement activation, lipid metabolism, angiogenesis, inflammation, and extracellular matrix remodeling \cite{MITCHELL20181147}. These interconnected mechanisms highlight the multifactorial nature of AMD and underscore the challenges in understanding and treating this condition effectively.
Globally, AMD constitutes 8.7\% of blindness cases and stands as a primary cause of blindness in developed nations, particularly affecting individuals aged 60 and above \cite{WONG2014e106}.

\vspace{0.05in}
\subsubsection{\textbf{Retinal Vessel Segmentation}}
Retinal vessel segmentation is crucial for extracting detailed information on the shape, thickness, and curvature of retinal blood vessels, offering valuable insights into various diseases. This technique is particularly significant in identifying conditions such as DR and macular degeneration and aiding in the early diagnosis of glaucoma by analyzing the blood vessel structure \cite{5660089, 7530915, SCHMIDTERFURTH20181}. Examination of critical features such as shape, orientation, width, curvature, branching patterns, and abnormal region volumes makes the blood vessel structure a pivotal source of essential information for disease analysis.
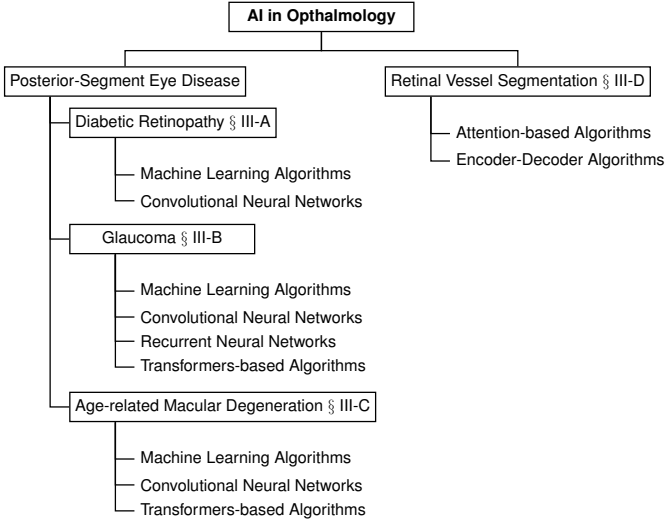
\begin{figure}[t]
    \centering
    \resizebox{\columnwidth}{!}{%
        \begin{forest}
            forked edges, folder indent=1cm,
            where={level()<1}{}{folder, grow'=east},
            where={level()>0}{l sep+=1cm}{},
            for tree={
                fork sep=4mm,
                thick, edge=thick,
                font=\sffamily,
                if n children=0{if n=1{yshift=-5mm}{}, for parent={s sep=0mm}}{draw, minimum height=4ex, minimum width=4cm, minimum width=4cm}
            }
            [AI in Opthalmology, calign=edge midpoint, s sep=2cm, font=\bfseries\sffamily
                [Posterior-Segment Eye Disease
                    [Diabetic Retinopathy $\S$ \ref{subsec:dr}
                        [Machine Learning Algorithms]
                        [Convolutional Neural Networks]
                    ]
                    [Glaucoma $\S$ \ref{subsec:glaucoma}
                        [Machine Learning Algorithms]
                        [Convolutional Neural Networks]
                        [Recurrent Neural Networks]
                        [Transformers-based Algorithms]
                    ]
                    [Age-related Macular Degeneration $\S$ \ref{subsec:AMD}
                        [Machine Learning Algorithms]
                        [Convolutional Neural Networks]
                        [Transformers-based Algorithms]
                    ]
                ]
                [Retinal Vessel Segmentation $\S$ \ref{subsec:vessel}
                    [Attention-based Algorithms]
                    [Encoder-Decoder Algorithms]
                ]
            ]
        \end{forest}%
    }
    \caption{Overview of our survey, highlighting application categories and key methods are used.}
    \vspace{-0.1in}
    \label{fig:taxonomy}
\end{figure}

\subsection{\textbf{Related Work}}
Several studies have explored recent advancements in AI methods for diagnosing eye diseases. These papers highlight the potential of modern approaches in ophthalmology, acknowledge existing challenges, and propose future research directions.

Litao et al. \cite{litao2021} review 143 papers, providing a structured framework for DL in ophthalmology, particularly for fundus images, and highlighting 33 publicly available datasets for early disease screening. Similarly, \cite{PAVITHRA2023157} explores the automated detection of diabetic macular edema using traditional and DL methods with retinal fundus and OCT images, detailing public datasets and the evolution of detection techniques. Ilesanmi et al. \cite{ILESANMI2023100261} focus on CNNs for retinal fundus image segmentation and classification, analyzing 62 studies and showcasing CNNs’ ability to enhance precision and achieve high accuracies with reduced reliance on human experts. Collectively, these works highlight the transformative role of DL, especially CNNs, in advancing retinal image analysis.

The review by \cite{LI2023101095} in another direction focuses on both posterior- and anterior-segment diseases. It highlights critical issues like real-world performance, generalizability, and interpretability, which require further attention. Additionally, \cite{9504555} focuses on retinal blood vessel segmentation using DL, exploring network architectures, trends, and challenges, while Xu et al. \cite{9745178} analyze enhancement techniques in DL for segmentation, drawing insights from 110 papers (2016-2021) to guide future improvements in accuracy and generalization.

Toward major eye disease prediction, Balla Goutam et al. \cite{9783206} provide a comprehensive review of DL strategies designed for DR, glaucoma, AMD, cataract, and retinopathy of prematurity. They examine the implementation pipeline, datasets, evaluation metrics, DL models and highlight eight key research directions for future progress in retinal disease diagnosis. Similarly, \cite{10.1007/s11042-023-15110-9} compares the efficacy of DL to traditional ML methods for DR diagnosis, emphasizing the need for collaboration with experts and further research to address remaining gaps in clinical settings.

Overall, these studies highlight significant advancements in AI-based methods for ophthalmic disease diagnosis. Building on this progress, we extend these efforts by exploring multiple ophthalmology-related applications, including DR, glaucoma detection, retinal vessel segmentation, and AMD. Additionally, we focus on cutting-edge DL architectures, such as transformers and multi-modal learning, to further enhance the capabilities of AI in ophthalmology (Figure \ref{fig:taxonomy}).

\section{\textbf{AI Application in Ophthalmology}}
\label{sec:applications}
\subsection{\textbf{Diabetic Retinopathy}}
\label{subsec:dr}

In this section, we explore recent advancements in AI applications for DR diagnosis, with a particular emphasis on improving interpretability. A summary of these algorithms is provided in Table \ref{tab:dr_ds_method}, with results derived from various datasets discussed in the respective papers.
\vspace{0.1in}
\subsubsection{\textbf{Machine Learning Algorithms}}
The work \cite{yagin2023explainable} focuses on using XAI to diagnose and treat DR in type 2 diabetes patients. ML models were created using clinical, biochemical, and metabolomic biomarkers to classify DR subclasses. Various ML techniques such as extreme gradient boosting (XGBoost) \cite{chen2016xgboost}, natural gradient boosting \cite{duan2020ngboost} for probabilistic prediction (NGBoost), and explainable boosting machine (EBM) \cite{nori2019interpretml} were compared for their performance.
\begin{figure}[t]
    \centering
    \includegraphics[width=0.48\textwidth]{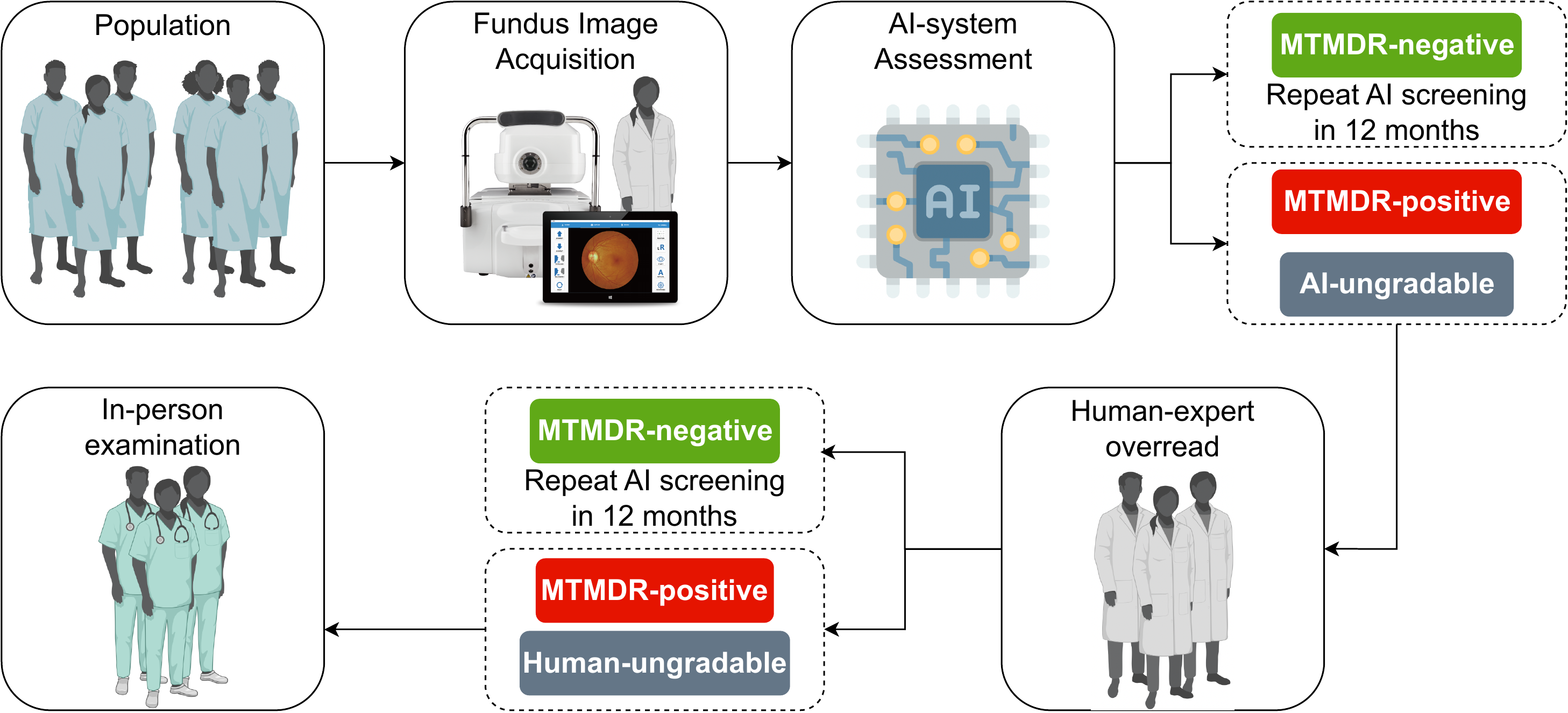}
    \caption{Proposed AI-human hybrid workflow: AI-screened fundus images labeled as more-than-mild diabetic retinopathy (MTMDR)-positive or AI-ungradable are overread by a human expert in teleophthalmology. Patients with an MTMDR-negative outcome undergo AI rescreening in 12 months, while those with an MTMDR-positive result or ungradable images are referred for in-person examination \cite{dow2023ai}.}
    \label{fig:dow2023ai}
\end{figure}

The study conducts feasibility analysis to pinpoint essential risk elements associated with the onset of DR in individuals with type 2 diabetes \cite{lalithadevi2023feasibility}. Employing a random forest (RF) model, the research achieves a robust prediction of DR prevalence, boasting an accuracy of 94.9\%. Model interpretation is facilitated through shapley additive explanations (SHAP) tools \cite{scott2017unified}. Another approach \cite{dow2023ai}, shown in Figure \ref{fig:dow2023ai}, proposes a method for DR detection that compares AI performance with human-based teleophthalmology. The study highlights the importance of making vision-preserving healthcare more accessible outside specialized eye care settings. To achieve this, the authors introduce an innovative AI-human hybrid workflow, where an AI algorithm conducts the initial assessment, which is then followed by overreading by retina specialists, significantly enhancing specificity while maintaining high sensitivity.

\vspace{0.1in}
\subsubsection{\textbf{Convolutional Neural Networks}}
In the study by Alghamdi et al. \cite{alghamdi2022towards}, an approach for explaining and validating model decisions in CNNs-based architectures is introduced using three DL models (VGG-16 \cite{simonyan2014very}, ResNet-18 \cite{he2016deep}, DenseNet-121 \cite{huang2017densely}) for diabetic retinopathy detection. The paper highlights the use of Grad-CAM \cite{selvaraju2017grad} visualizations to assess model interpretability, revealing the superiority of the VGG-16 model. Similarly, Boreiko et al. \cite{boreiko2022visual} explore DL model interpretability for diabetic retinopathy detection, presenting an ensemble approach that combines plain and adversarially robust models. This ensemble not only improves accuracy but also enhances visual explanations with meaningful visual counterfactual explanations (VCEs).

Che et al. \cite{che2022learning} propose a framework for joint grading of DR and diabetic macular edema, using dynamic difficulty-aware weighted loss (DAW) and a dual-stream disentangled learning architecture (DETACH). DAW adapts the difficulty-aware parameter to help the model manage challenging cases, while DETACH improves interpretability and performance by independently focusing on specific aspects of each pathology. Similarly, DRG-Net \cite{tusfiqur2022drg} addresses both disease grading and multi-lesion segmentation, automatically selecting the most relevant lesion information while offering explainable properties.

Hesse et al. \cite{hesse2022insightr} introduce INSightR-Net, an interpretable CNN designed for DR grading. This model incorporates a prototype layer to visualize image areas similar to learned prototypes, framing the final prediction as a weighted average of prototype labels based on similarity. INSightR-Net achieves competitive performance compared to a ResNet baseline, demonstrating that interpretability can be attained without sacrificing accuracy. Building on this, Jiang et al. \cite{jiang2023eye} focus on improving early DR detection by combining eye-tracking technology with DL models. Their multi-modal approach integrates gaze maps, captured during fundus image examinations, into a DL architecture using attention mechanisms. This gaze map attention-guide model enhances both accuracy and interpretability, offering a robust and generalizable solution for DR detection.

\begin{figure}[H]
    \centering
    \includegraphics[width=0.5\textwidth]{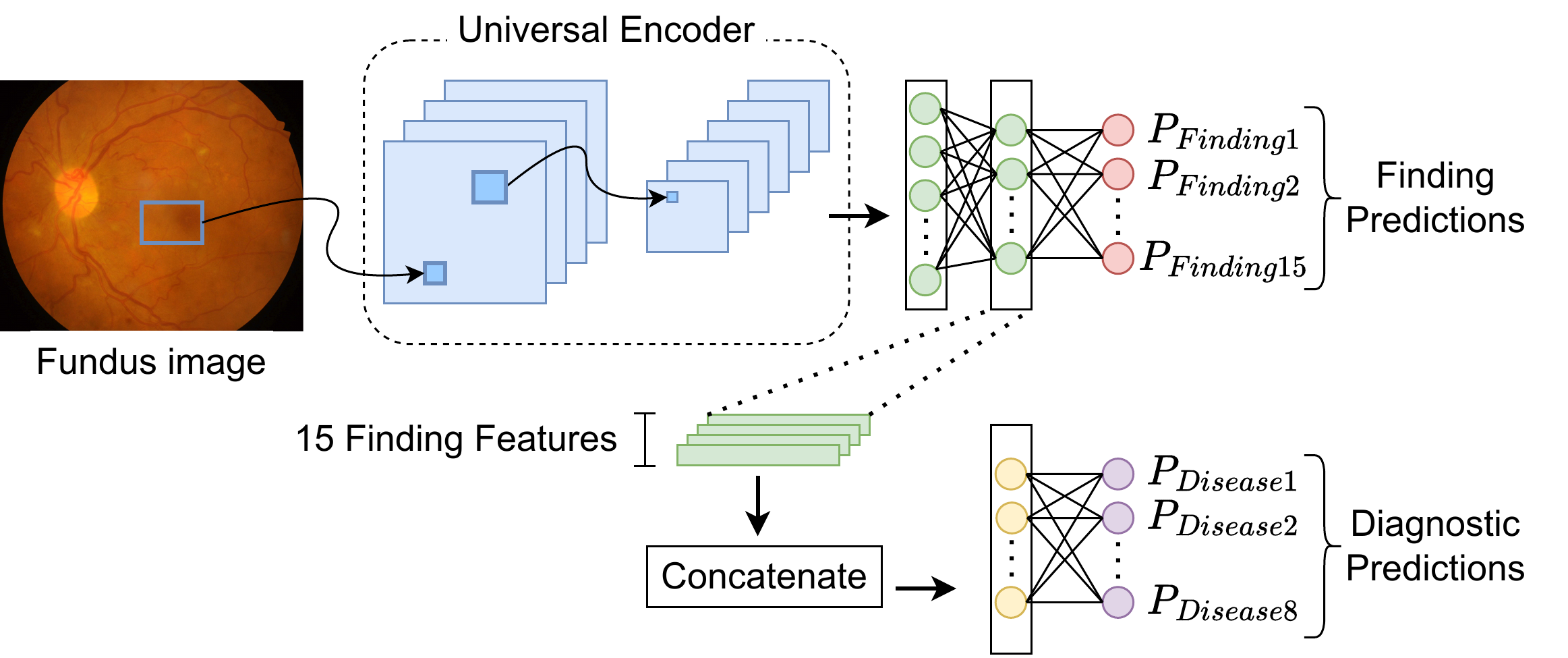}
    \caption{Overall architecture of deep learning-based Computer-Aided Diagnosis for diabetic retinopathy detection
    \cite{son2023interpretable}.
    }
    \label{fig:son2023interpretable}
\end{figure}

In the work by Son et al. \cite{son2023interpretable} as shown in Figure \ref{fig:son2023interpretable}, a DL-based computer-aided diagnosis (CAD) system is introduced for comprehensive retinal analysis, identifying 15 abnormal retinal findings and diagnosing eight major ophthalmic diseases. The system emphasizes interpretability through the counterfactual attribution ratio (CAR), providing a transparent diagnostic reasoning process.

\begin{table*}[t]
\begin{center}
\caption{Typical methods on the application of AI in diabetic retinopathy.}
\label{tab:dr_ds_method}
\setlength{\tabcolsep}{3pt}
\scalebox{0.8}{
\begin{tabular}{c| llp{60pt} p{240pt}}
\toprule
\textbf{Year} &  \textbf{Study} & \textbf{Methodology} & \textbf{Task} & \textbf{Performance} \\
\toprule
2021 & Che Haoxuan et al. \cite{che2022learning} & Joint feature representation & DR Grading & AUC=0.797, F1=0.308, Acc=0.429, Rec=0.370, Pre=0.365\\ \midrule
\multirow{4}{*}{2022} & Alghamdi Hanan Saleh \cite{alghamdi2022towards} & VGG-16, ResNet-18, DenseNet-121 & DR Detection & APTOS: Precision VGG16=0.87, ResNet18=0.67, DenseNet121=0.74 \\ \cmidrule{2-5}
& Boreiko Valentyn et al. \cite{boreiko2022visual}  & DNN & DR Detection & Acc=0.89\\ \cmidrule{2-5}
& Obayya Marwa et al. \cite{obayya2022explainable}  & UNet, SqueezeNet & DR Detection & Acc= 0.96 - 0.98\\ \cmidrule{2-5}
& Hesse Linde et al. \cite{hesse2022insightr} & INSightR-Net & DR Grading & MAE=0.59 \\ \midrule
\multirow{5}{*}{2023} & Fatma Hilal Yagin et al. \cite{yagin2023explainable} & XGBoost, NGBoost & DR Detection & XGBoost: Acc=0.913, Pre=0.893, Rec=0.912, F1=0.894, AUROC=0.97; NGBoost: Acc=0.881, Pre=0.881, Rec=0.881, F1=0.881, AUROC=0.96\\ \cmidrule{2-5}
& Jiang Hongyang et al. \cite{jiang2023eye}  & Supervised mask guides DNN attention. & DR Detection & Sen=0.783, Spec=0.683, Acc = 0.733, F1 = 0.746\\ \cmidrule{2-5}
& Lalithadevi Balakrishnan et al. \cite{lalithadevi2023feasibility}  & Random Forest & DR Detection & Acc=0.949\\ \cmidrule{2-5}
& Eliot R. Dow et al. \cite{dow2023ai} &  - & DR Detection & Sen=0.95 \\ \cmidrule{2-5}
& Son Jaemin et al. \cite{son2023interpretable}  & DNNs & Multi-diseases classification & AUROC=0.992\\  
\cmidrule{2-5}
& Tusfiqur, Hasan Md, et al. \cite{tusfiqur2022drg} & Joint Learning of DR lesion and classification & DR Grading,  DR Detection & Acc = 0.87 - 0.94\\ 
\bottomrule
\end{tabular}
}
\end{center}
\end{table*}
\vspace{0.1in}
\subsubsection{\textbf{Datasets}}
Below, we provide a summary of the most commonly used datasets for training and evaluating DL models in the context of diabetic retinopathy-related diseases with a comparison presented in Table \ref{tab:dr_ds_compare}.

\paragraph{DIARETDB0 and DIARETDB1} The DIARETDB0  \cite{Kauppi2007TheDD} database includes 130 retinal images captured with a digital fundus camera (50 degrees FOV), comprising 20 healthy images and the rest exhibiting DR symptoms. DIARETDB1 \cite{Kauppi2007TheDD}, with 89 retinal fundus images, predominantly contains mild non-proliferative diabetic retinopathy (NPDR) signs (84 images), along with 5 healthy ones. Both databases share a resolution of 1500$\times$1152 pixels.
\paragraph{Retinopathy Online Challenge} The Retinopathy Online Challenge (ROC) \cite{Niemeijer2010} microaneurysms database was created for an online competition aimed at developing the most effective algorithm for identifying microaneurysms in retinal images. The images were captured using Topcon NW 100, Topcon NW 200, or a Canon CR5-45NM, saved in JPEG compression format.

\paragraph{Messidor} The Messidor \cite{10.1001/jamaophthalmol.2013.1743} database, designed for evaluating segmentation and indexing techniques in retinal ophthalmology, comprises 1200 retinal fundus images captured using a 3CCD video camera on a Topcon TRCNW6 nonmydriatic retinography with a 45-degree field of view. The images come in three resolutions: 1140$\times$960, 2240$\times$1488, and 2304$\times$1536 pixels.

\paragraph{e-ophtha EX and e-ophtha MA} The database \cite{Decencière2013} contains 82 retinal images, consisting of 47 pathological and 35 nonpathological images. Captured at varying resolutions from the OPHIDAT medicine center, the images feature diverse sizes and shapes of exudates. A subset of E-Ophtha, is designed for the study of microaneurysms, featuring 381 images. Among them, 148 images exhibit small or large microaneurysms, while 233 images are normal.

\paragraph{Messidor-2} The Messidor-2   \cite{Decencière2014} dataset focuses on DR examinations, presenting pairs of macula-centered eye fundus images for each examination. The Messidor-Original subset comprises 529 examinations (1058 images in PNG format) from the original Messidor dataset. The Messidor-Extension subset adds 345 examinations (690 images in JPG format). In total, Messidor-2 encompasses 874 examinations (1748 images) and is accompanied by a spreadsheet detailing image pairing.

\paragraph{RC-RGB-MA} The RC-RGB-MA  \cite{RetinaCheckTeam2016} dataset is part of the RetinaCheck project led by Eindhoven University of Technology, the Netherlands. It consists of 250 RGB retinal images captured using a diabetic retinopathy severity non-mydriatic fundus camera.

\paragraph{IDRiD} IDRiD \cite{h25w98-18} is a retinal image dataset designed for evaluating algorithms in the automatic detection and grading of DR and Macular Edema. It includes 516 images with marked OD center and fovea and 81 images with segmented optic disc boundaries. Acquired using a Kowa VX-10 alpha digital fundus camera, the images have a resolution of 4288$\times$2848 pixels and a 50-degree field of view.

\paragraph{DDR} The study \cite{LI2019} gathered 13,673 fundus images from 9,598 patients to assess various methods in clinical settings. Seven graders categorized the images into six classes based on image quality and DR level. Additionally, 757 images with DR were chosen for annotation, focusing on four types of DR-related lesions. 

\paragraph{Kaggle DR} The Kaggle DR \cite{diabetic-retinopathy-detection} dataset is the largest collection of fundus images  with 88,702 samples utilized for DR classification.

\paragraph{APTOS} The APTOS 2019 Blindness Detection (APTOS 2019 BD) \cite{aptos2019-blindness-detection} dataset comprises 3,662 fundus photographs collected from rural India by the Aravind Eye Hospital. Trained doctors labeled images based on the International Clinical Diabetic Retinopathy Severity Scale
(ICDRSS), resulting in five categories: no DR, mild DR, moderate DR, severe DR, and proliferative DR.

\paragraph{DeepDRiD} The ``Diabetic Retinopathy (DR)-Grading and Image Quality Estimation Challenge'' \cite{LIU2022100512} conducted in collaboration with ISBI 2020 encompassed three sub-challenges focused on developing DL models for DR image assessment and grading. 



\begin{table}[!hbt]
\begin{center}
\caption{List of datasets supporting Diabetic Retinopathy.}
\vspace{0.05in}
\label{tab:dr_ds_compare}
\setlength{\tabcolsep}{3pt}
\scalebox{0.8}{
\begin{tabular}{l|lccc}
\toprule
\textbf{Year} & \textbf{Dataset} & \textbf{\# Images} & \textbf{Format} & \textbf{Resolution} \\
\toprule
\multirow{2}{*}{2007} & DIARETDB0 \cite{Kauppi2007TheDD} &  130 & JPEG/PNG & 1500$\times$1152 \\ \cmidrule{2-5}
& DIARETDB1 \cite{Kauppi2007TheDD} & 89 & JPEG/PNG & 1500$\times$1152 \\ \midrule
2010 & ROC \cite{Niemeijer2010} & 100 & JPEG & various\\ \midrule
\multirow{3}{*}{2013}  & Messidor \cite{10.1001/jamaophthalmol.2013.1743} & 1200 & JPEG & various \\ \cmidrule{2-5}
&  e-ophtha EX \cite{Decencière2013} &  82 &-& various \\ \cmidrule{2-5}
& RC-RGB-MA \cite{RetinaCheckTeam2016} &  250 & JPEG & 2595$\times$1944 \\ \cmidrule{2-5}
& e-ophtha MA \cite{Decencière2013} & 381 &-& various \\ \midrule
2014
& Messidor-2 \cite{Decencière2014} &  1748 & JPEG & various\\ \midrule
2015 & Kaggle DR \cite{diabetic-retinopathy-detection} & 88,702 & JPEG & various \\ \midrule
2018 & IDRiD \cite{h25w98-18} & 516 & JPEG & 4288$\times$2848 \\ \midrule
\multirow{2}{*}{2019} & DDR \cite{LI2019} & 13,673 &-& various \\ \cmidrule{2-5}
& APTOS \cite{aptos2019-blindness-detection}&13,000&PNG &-\\ \midrule
2022 & DeepDRiD \cite{LIU2022100512}&2,000&-&various\\ \bottomrule
\end{tabular}
}
\label{tab1}
\end{center}
\end{table}

\subsection{\textbf{Glaucoma}}
\label{subsec:glaucoma}
Modern methodologies have emerged as impactful tools in automated glaucoma detection, progression prediction, and the segmentation of critical anatomical structures. This section delves into recent advancements in AI-driven approaches for glaucoma analysis, showcasing innovative techniques and their contributions to enhancing clinical accuracy, efficiency, and outcomes. Table \ref{tab:glaucoma_method} provides a comprehensive comparison of these methods, highlighting their core methodologies, reported performance metrics, and the datasets utilized, offering valuable insights into the evolving landscape of AI applications in glaucoma care.

\vspace{0.1in}
\subsubsection{\textbf{Machine Learning Algorithms}} 
The study by \cite{deperlioglu2022explainable} focuses on developing a ML model for glaucoma diagnosis, highlighting the exceptional performance of the XGBoost algorithm, which achieved an accuracy of 0.947. This work enhances the XAI literature by demonstrating the efficacy of a hybrid system for diagnosing glaucoma. Similarly, \cite{oh2021explainable} presents a glaucoma prediction model incorporating an embedded explanation system to improve interpretability. Using comprehensive clinical data, including visual field tests, retinal nerve fiber layer (RNFL) OCT tests, general examinations, and fundus image tests, the authors employed multiple algorithms --- support vector machine \cite{cortes1995support,cristianini2000introduction}, C5.0 \cite{quinlan2014c4,kuhn2013applied}, random forest \cite{breiman2001random}, and XGBoost \cite{chen2016xgboost} --- with XGBoost again proving most effective. To provide insights into the decision-making process, the study utilized interpretability tools such as gauge and radar charts alongside shapley additive explanations analysis \cite{scott2017unified}, offering a clear view of individual predictions.
\vspace{0.1in}
\subsubsection{\textbf{Convolutional Neural Networks}} 

\begin{figure}[!hbt]
    \centering
    \includegraphics[width=0.45\textwidth]{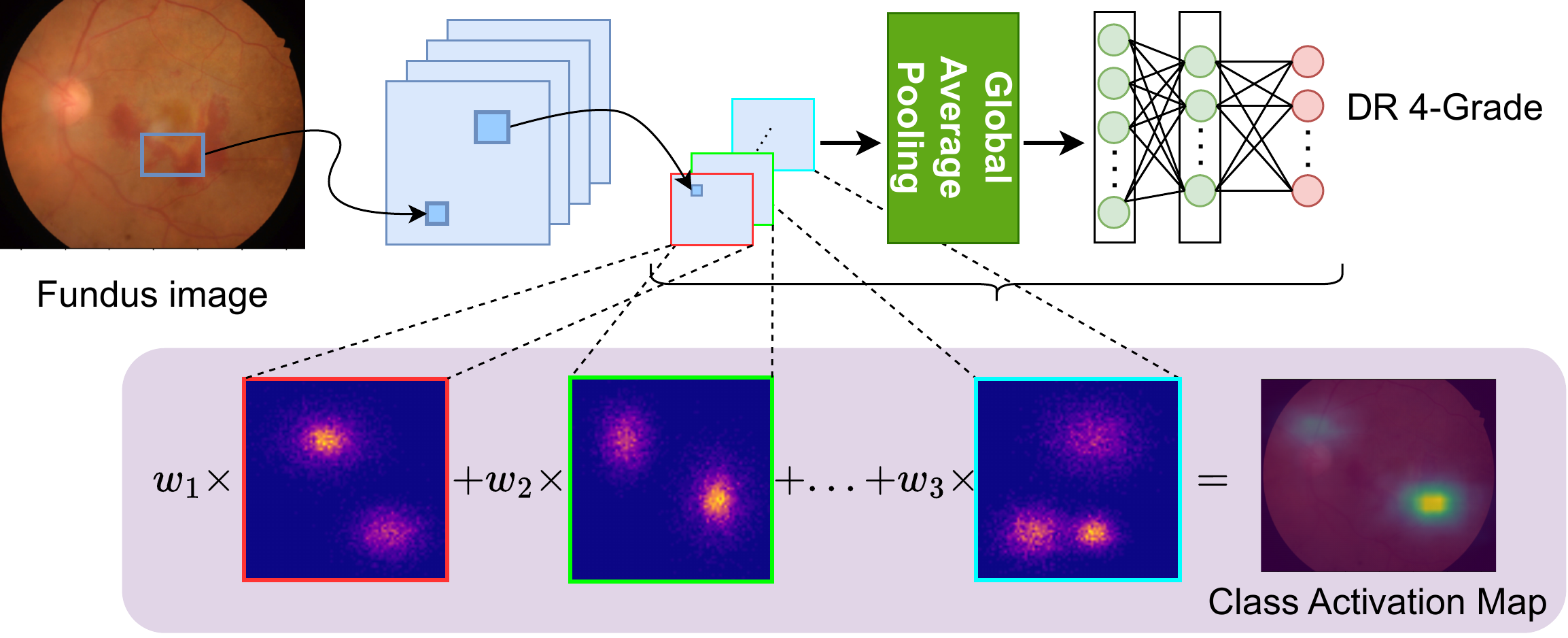}
    \caption{CAM: Class Activation Mapping for ensuring explainability in CNN models \cite{deperlioglu2022explainable,selvaraju2017grad}.}
    \label{fig:deperlioglu2022explainable}
\end{figure}

The reviewed studies collectively advance the application of XAI and CNNs-based models for glaucoma diagnosis and related tasks, offering diverse methodologies and valuable insights. In \cite{chang2021explaining}, adversarial examples and GradCAM are employed to interpret DL models trained on retinal fundus images for glaucoma detection, with specialists favoring adversarial examples for superior explainability. Similarly, \cite{deperlioglu2022explainable} introduces a hybrid CNN-based solution supported by CAM visualizations, highlighting the integration of DL and image processing for glaucoma diagnosis as illustrated in Figure \ref{fig:deperlioglu2022explainable}.

Building on these efforts, \cite{hemelings2021deep} presents a ResNet-based framework for glaucoma detection and vertical cup-to-disc ratio (VCDR) estimation, emphasizing the importance of optic disc localization while achieving robust performance even on challenging image datasets. Further, \cite{kamal2022explainable} proposes AFFD-Net for retinal vessel segmentation, outperforming state-of-the-art methods, and introduces a two-step glaucoma prediction process combining spike neural networks (SNNs) and adaptive neuro-fuzzy inference systems (ANFIS), with strong interpretability through XAI and interpretable machine learning (IML) techniques. Meanwhile, \cite{sigut2023interpretable} employs decision trees \cite{shalev2014understanding} as surrogate models to enhance the interpretability of CNN predictions, achieving consistent results across architectures.

Other studies investigate model-specific contributions. Akter et al. \cite{akter2023glaucoma} compare SqueezeNet \cite{iandola2016squeezenet}, ResNet18 \cite{he2016deep}, and VGG16 \cite{simonyan2014very} for glaucoma detection from raw OCT scans, with VGG16 delivering the most effective performance. Similarly, Fan et al. \cite{fan2022detecting} evaluate a ResNet-50 model trained on fundus photographs from the OHTS dataset, achieving strong diagnostic precision. The study by Schottenhamml et al. \cite{schottenhamml2023impact} also highlights the superiority of CNNs over traditional vessel density biomarkers for glaucoma detection. Their findings demonstrate the robustness of CNNs across varying scan sizes, reinforcing their applicability in diverse clinical settings. Meanwhile, M. Yan et al. \cite{yan2023mix} address the critical challenge of domain gaps in glaucoma detection within fundus images. By combining domain adaptation and domain mixup techniques, their approach effectively mitigates discrepancies between datasets, leading to enhanced generalization performance. This novel strategy underscores the potential of leveraging domain adaptation to improve model reliability across heterogeneous datasets.

Recent advancements also extend beyond fundus imaging. Braeu et al. \cite{braeu2023geometric} leverage geometric DL techniques \cite{bronstein2017geometric,monti2017geometric}, comparing DGCNN \cite{wang2019dynamic} and PointNet \cite{qi2017pointnet} for glaucoma detection from 3D optic nerve head (ONH) point clouds, effectively analyzing critical structural features. Agboola et al. \cite{agboola2023wavelet} explore wavelet scattering networks (WSNs) for feature extraction from fundus images, demonstrating high accuracy for glaucoma detection without traditional preprocessing steps.


In progression analysis, Mariottoni et al. \cite{mariottoni2023deep} propose a deep network utilizing RNFL thickness measurements from SD-OCT scans to predict glaucoma progression, outperforming traditional trend-based analyses and altering post-test probability estimates with interval likelihood ratios. Fei Li et al. \cite{li2022deep} present a DiagnoseNet network, which comprises a segmentation module using U-Net \cite{ronneberger2015u} to extract four anatomical structures (retinal vessels, macula, optic cup, and optic disk), which are merged into a single channel to enhance focus. This augmented image is then processed by the diagnostic module, built on EfficientNet-B0 \cite{tan2019efficientnet}, with modifications to fine-tune it for binary classification.

Innovative glaucoma-related segmentation approaches are also highlighted by Meng et al. \cite{meng2022shape}, Shyla et al. \cite{shyla2023glaucoma}, and Mangipudi et al. \cite{mangipudi2021improved}. For instance, Meng et al. \cite{meng2022shape} introduce a weakly supervised framework leveraging a modified signed distance function (SDF) and dual consistency regularization for spatially aware optic disc and cup segmentation, coupled with an end-to-end vertical cup-to-disc ratio estimation method. Shyla et al. \cite{shyla2023glaucoma} combine level set segmentation with AlexNet \cite{krizhevsky2012imagenet} classification, emphasizing the importance of integrating DL with advanced segmentation techniques. Mangipudi et al. \cite{mangipudi2021improved} in other direction develop a robust DL system for optic disc and cup segmentation, employing probabilistic ground truth masks and tailored loss functions to manage uncertainty and enhance model performance.

These works, in general, jointly underscore the integration of explainable and interpretable AI techniques across diverse data modalities and tasks in glaucoma research, paving the way for more effective and transparent clinical applications.

\begin{table*}[t]
\begin{center}
\caption{Typical studies on the application of AI in Glaucoma.}
\label{tab:glaucoma_method}
\setlength{\tabcolsep}{3pt}
\scalebox{0.8}{
\begin{tabular}{l|p{70pt} p{100pt} p{90pt} p{330pt}}
\toprule
\textbf{Year} & \textbf{Study} & \textbf{Method} & \textbf{Task} & \textbf{Performance} \\
\toprule
\multirow{5}{*}{2021}
& Ruben et al. \cite{hemelings2021deep} & ResNet-50 & Glaucoma Detection & AUC=0.94\\ \cmidrule{2-5}
& Sejong et al. \cite{oh2021explainable}& SVM, C50, Random Forest, and XGBoost& Glaucoma Classification &SVM:Acc=0.925, Sen=0.933, Spec=0.920, AUC=0.945; C50: Acc=0.903, Sen=0.874, Spec=0.92, AUC=0.897; Random Forest: Acc=0.937, Sen=0.924, Spec=0.945, AUC=0.945; XGBoost: Acc=0.947, Sen=0.941, Spec=0.950, AUC=0.945\\ \cmidrule{2-5}
& Sarathi et al. \cite{mangipudi2021improved} & Salient Point Detection, CNNs& Disc \& Cup Segmentation & DRISHTI: IoU\_disc=0.966, Dice\_disc=0.9529; IoU\_cup=0.944, Dice\_cup=0.933; RIM-ONE: IoU\_disc=0.9842, Dice\_disc=0.9452; IoU\_cup=0.6592 , Dice\_cup=0.7863; DRIONS-DB:  IoU\_disc=0.9615, Dice\_disc=0.9547\\ \cmidrule{2-5}
&Jooyoung et al. \cite{chang2021explaining}&DNNs& Glaucoma Classification&AUC=0.90, Sen=0.79\\ \midrule

\multirow{5}{*}{2022} &  
Junde et al. \cite{wu2022learning}&RNN for OD/OC segmentation calibration& Disc \& Cup Segmentation & Optic Disc: Dice=0.963; Optic Cup: Dice=0.897\\ \cmidrule{2-5}
& Rui et al. \cite{fan2022detecting} & ResNet-50& Glaucoma Detection & AUROC=0.88 \\  \cmidrule{2-5}
& Sarwar et al. \cite{kamal2022explainable} & Spike Network & Glaucoma Detection & Pre=0.96, Rec=0.95, ACC=0.96, F-score=0.97\\  \cmidrule{2-5}
&Omer et al. \cite{deperlioglu2022explainable}  & CNN & Glaucoma Detection & ACC=0.93, Rec=0.97, AUC=0.95, F1=0.95, Pre=0.93\\ \cmidrule{2-5}
& Yanda et al. \cite{meng2022shape} & Weakly/semi-supervised framework & Disc \& Cup Segmentation & Optic Disc: Dice=0.871; Optic Cup: Dice=0.972\\ \midrule

\multirow{14}{*}{2023} & 
Hyla et al. \cite{shyla2023glaucoma}& AlexNet & Glaucoma Classification & Acc=0.98, Sen=0.97, Spec=0.97\\ \cmidrule{2-5}
& Mariottoni et al. \cite{mariottoni2023deep}  & CNN & Glaucoma Progression & AUC=0.938, Sen=0.873, Spec=0.864\\ \cmidrule{2-5}
& Ming et al. \cite{yan2023mix} & Domain adaptation & Glaucoma Detection & Acc=0.967, Sen=0.955, Spec=0.969, AUC=0.995\\ \cmidrule{2-5}
& Hafeez et al. \cite{agboola2023wavelet} & Wavelet Scattering Network & Glaucoma Detection & Acc=0.98\\ \cmidrule{2-5}
& Rui et al. \cite{fan2023detecting} & Transformers: DeiT & Glaucoma Detection & AUROC=0.91\\ \cmidrule{2-5}
& You et al. \cite{zhou2023representation} & Transformers & Glaucoma Classification & Kappa = 0.85, F1 = 0.91 (private test set), ACU = 0.99 (private test set) \\ \cmidrule{2-5}
& D. Leite et al. \cite{leite2023vision} & Transformers: ViT-BRSET & Glaucoma (optic nerve excavation) Detection  & Acc = 0.94, F1 = 0.91, Recall = 0.94 \\ 
\cmidrule{2-5}
& Shaista et al. \cite{hussain2023predicting} & LSTM \& CNN Combination & Glaucoma Progression & AUC=0.83\\ \cmidrule{2-5}
&Vutukuru et al. \cite{kumar2023novel}& Segmentation via Unet++ and ResNet with GRU optimization&Glaucoma Detection&Acc=0.988, Sen=0.992, Spec=0.983\\ \cmidrule{2-5}

& Jose et al. \cite{sigut2023interpretable} &VGG19, ResNet50, InceptionV3, Xception, and Decision Tree Classifier &Glaucoma Classification&VGG19=0.901, ResNet50=0.899, InceptionV3=0.904, Xception=0.897\\ \cmidrule{2-5}

&Fabian A et al. \cite{braeu2023geometric} & Dynamic Graph CNNs (DGCNN), PointNet & Glaucoma Detection &DGCNN: AUC=0.97, PointNet: AUC=0.95\\ \cmidrule{2-5}

&Nahida et al. \cite{akter2023glaucoma}& SqueezeNet, ResNet18, VGG16&Glaucoma Detection&SqueezeNet: AUC=0.973, Acc=0.936, Sen=0.945, Spec=0.927, Pre=0.928, F1=0.936; ResNet-18: AUC=0.978, Acc=0.94, Sen=0.936, Spec=0.945, Pre=0.944, F1=0.94; VGG16: AUC=0.988, Acc=0.952, Sen=0.945, Spec=0.959, Pre=0.958, F1=0.95\\ \cmidrule{2-5}
 
& Julia et al. \cite{schottenhamml2023impact} &  DenseNets & Glaucoma Detection & AUROC=0.89 (3x3 mm macular), AUROC=0.93 (6x6 mm macular), AUROC=0.89 (6x6 mm ONH scans)\\ \cmidrule{2-5}

&Ming et al. \cite{yan2023mix} &   ResNeST, Domain Adaptation & Glaucoma Detection & REFUGE: Sen=0.875, AUC=0.990; LAG: Acc=0.967, Sen=0.956, Spec=0.969, AUC=0.995; ORIGA: Sen=0.698, Spec=0.852, AUC=0.891; RIM-ONE: Acc=0.954, Sen=0.920, Spec=0.980, AUC=0.993\\ \bottomrule
\end{tabular}
}
\label{tab1}
\end{center}
\end{table*}

\vspace{0.1in}
\subsubsection{\textbf{Recurrent Neural Networks}}
\begin{figure}[t]
    \centering
    \includegraphics[width=0.5\textwidth]{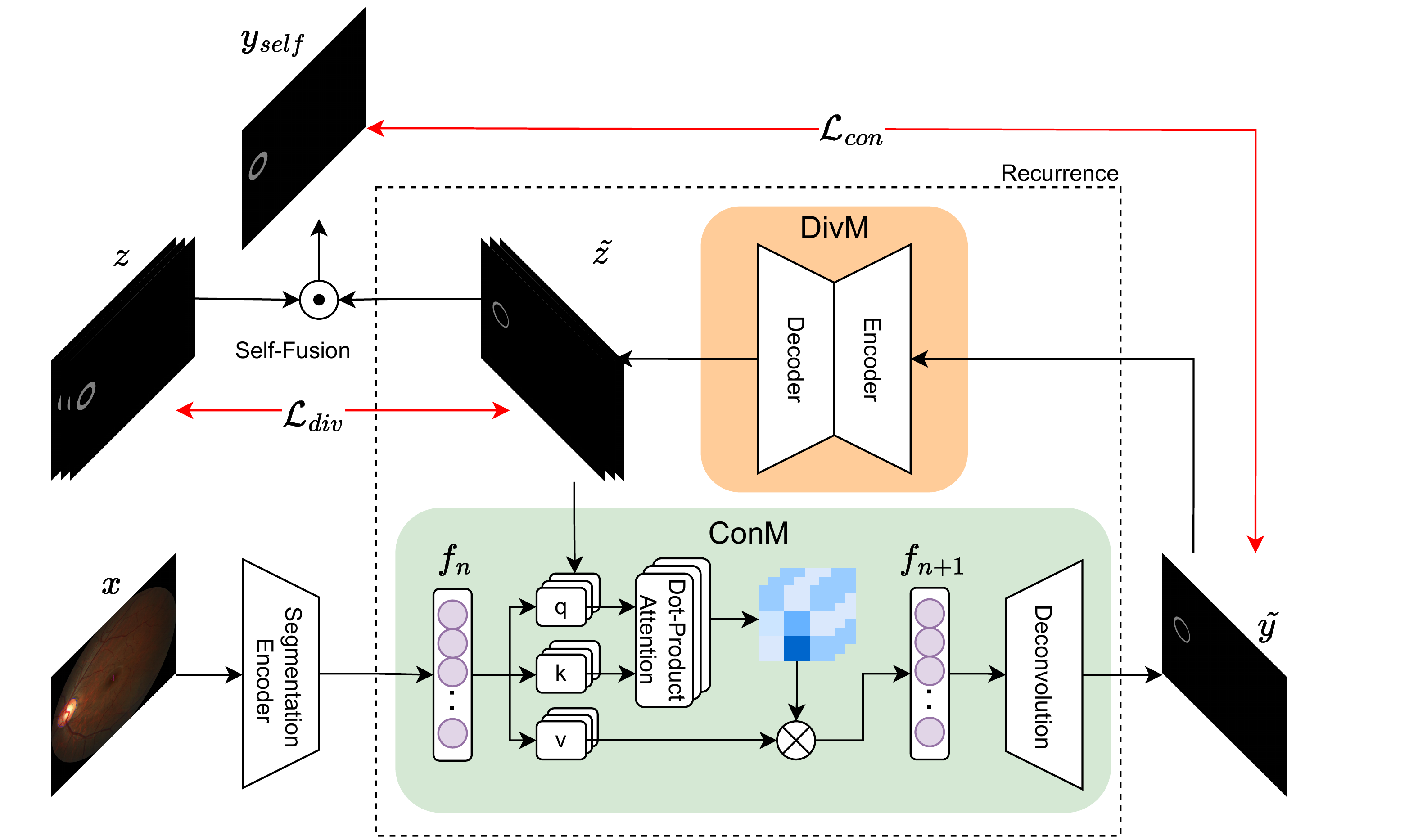}
    \caption{Overall architecture of the proposed self-calibration segmentation method for glaucoma-related disease. Green denotes ConM modules. Orange denotes DivM modules \cite{wu2022learning}.}
    \label{fig:wu2022learning}
\end{figure}

The research in \cite{wu2022learning} presents an innovative recurrent neural network framework tailored for glaucoma diagnosis, specifically designed to tackle challenges linked with multi-rater annotations. The methodology is centered around a recurrent model proficient in learning self-calibrated segmentation from these annotations. In Figure \ref{fig:wu2022learning}, this intricate process incorporates both ConM (Calibration Module) and DivM (Division Module) within an iterative optimization framework. 


The study by Hussain et al. \cite{hussain2023predicting} introduces a multi-modal DL model that combines the LSTM network \cite{hochreiter1997long} with CNNs for predicting glaucoma progression. By integrating diverse data sources, including OCT images, visual field (VF) values, and demographic and clinical information, the model achieves high accuracy in forecasting VF changes up to 12 months in advance. A standout feature is the use of synthetic future images generated by a generative adversarial network (GAN) \cite{goodfellow2020generative}, which significantly enhances prediction performance.

In a related study, Kumar et al. \cite{kumar2023novel} propose a glaucoma detection framework that combines a gated recurrent unit (GRU)-based optimization \cite{cho2014learning} with the Unet++ architecture \cite{zhou2019unet++} and ResNet. Their approach begins with pre-processing retinal fundus images using histogram equalization, followed by disc and cup segmentation using the U-shape network \cite{ronneberger2015u,zhou2019unet++}, optimizing the accuracy of glaucoma detection.

\vspace{0.1in}
\subsubsection{\textbf{Transformers-based Algorithms}} The research presented in \cite{fan2023detecting} conducts a comparative analysis of the diagnostic accuracy and interpretability of ViT models \cite{dosovitskiy2020image} and ResNet-50 in distinguishing primary open-angle glaucoma from fundus photographs. The study evaluates diagnostic performance through metrics such as AUROC and sensitivity at fixed specificities, providing a thorough assessment of each model’s ability to detect glaucoma. The results highlight the advantages of ViTs, not only in enhancing generalization across diverse datasets but also in improving the interpretability of DL models for detecting eye diseases and other medical conditions. These findings align with similar studies \cite{bowd2022primary,leite2023vision}, which also emphasize the potential of ViTs in advancing both diagnostic performance and model transparency.

In related work, \cite{zhou2023representation} introduces MM-RAF, a transformer-based framework designed for multi-modal glaucoma recognition. MM-RAF effectively manages the cross-modality interactions between Color Fundus Photography (CFP) and OCT by incorporating specialized modules such as bilateral contrastive alignment, multiple instance learning representation, and hierarchical attention fusion. This innovative approach improves the integration of diverse imaging modalities, facilitating more accurate and robust glaucoma diagnosis.


\vspace{0.1in}
\subsubsection{\textbf{Datasets}}
\paragraph{ONHSD} The ONHSD \cite{ONHSD2012} (Optic Nerve Head Segmentation Database) includes 99 fundus images with a resolution of 640$\times$480 obtained from 50 patients selected randomly from a DR screening program. Among them, 90 images are allocated for evaluating segmentation algorithms, and 96 images feature a visible ONH. The images were captured using a Canon CR6 45MNf fundus camera with a 45-degree field angle lens.

\paragraph{Drions-DB} The Drions-DB \cite{Carmona:2008:ION:1383660.1383874} (Digital Retinal Images for Optic Nerve Segmentation Database) consists of 110 fundus images with a resolution of 600$\times$400 pixels. Approximately 23\% of the patients had chronic glaucoma, and 77\% had ocular hypertension. 

\paragraph{ORIGA} The ORIGA \cite{5626137} (Online Retinal Fundus Image Database for Glaucoma Analysis and Research) database comprises 650 images with optic disc (OD) and optic cup (OC) segmentation, along with information on glaucoma severity grading. However, it is not publicly available.

\paragraph{RIM-ONE} The Retinal Image Database for Optic Nerve Evaluation (RIM-ONE) comprises 169 images classified into different categories: 118 as normal, 12 as early glaucoma, 14 as moderate glaucoma, 14 as deep glaucoma, and 11 as ocular hypertension.

\paragraph{ACHIKO-K} The ACHIKO-K \cite{6566371} database consists of 258 manually annotated retinal images taken from glaucoma patients. The images include detailed information on glaucoma-related pathological signs such as hemorrhage, optic nerve drusen, and optic cup notching.

\paragraph{Drishti-GS} The Drishti-GS \cite{Sivaswamy2015ACR} database comprises 101 fundus images of the Indian population with a resolution of 2896$\times$1944 pixels. The training subset includes 50 images with optic disc (OD) and optic cup (OC) segmentation ground truths, along with notching information.

\paragraph{RIGA} The Retinal fundus images for glaucoma analysis (RIGA) \cite{Almazroa2018} dataset includes 750 fundus images with optic disc (OD) and optic cup (OC) segmentation ground truth.

\paragraph{LAG} The LAG \cite{Li_2019_CVPR} database comprises 5,824 fundus images, including 2,392 positive and 3,432 negative glaucoma samples obtained from Beijing Tongren Hospital.

\paragraph{REFUGE} The REFUGE \cite{ORLANDO2020101570} database comprises 1200 retinal images obtained from subjects of Chinese ethnicity using two devices: a Zeiss Visucam 500 fundus camera with a resolution of 2124$\times$2056 pixels (400 images) and a Canon CR-2 device with a resolution of 1634$\times$1634 pixels (800 images). 

\paragraph{Rotterdam EyePACS AIROGS} The Artificial Intelligence for Robust Glaucoma Screening (AIROGS) \cite{AIROGS2023} challenge aims to develop algorithms capable of robust glaucoma screening. It features a substantial dataset comprising approximately 113,000 images from over 60,000 patients across 500 screening centers.

\paragraph{EyePACS-AIROGS-light} The EyePACS-AIROGS-light \cite{10270429} dataset is derived from a balanced subset of standardized fundus images from the Rotterdam EyePACS AIROGS \cite{AIROGS2023} train set. It includes separate folders for training, validation, and testing, each containing a specific number of fundus images for referable glaucoma (RG) and non-referable glaucoma (NRG) classes.

\paragraph{EyePACS-AIROGS-light-v2} EyePACS-AIROGS-light-v2, as described in \cite{10.1145/3603765.3603779}, is similarly sourced from a well-proportioned selection of standardized fundus images found within the Rotterdam EyePACS AIROGS dataset \cite{AIROGS2023}. It consists of training, validation, and test folders, with approximately 84\%, 8\%, and 8\% of the total 4800 images allocated to each, respectively. Each class, referable glaucoma (RG) and non-referable glaucoma (NRG), has its own folder within the training set.

\paragraph{Chákṣu} The Chákṣu \cite{Kumar2023} database is designed for assessing computer-assisted glaucoma prescreening techniques, offering 1345 color fundus images captured with three different commercially available fundus cameras. It stands out as the largest Indian-ethnicity-specific fundus image database, featuring expert annotations.

\paragraph{GAMMA} The GAMMA dataset \cite{wu2023gamma} is developed by the Sun Yat-sen Ophthalmic Center at Sun Yat-sen University in Guangzhou, China, represents the world’s first multi-modal dataset for glaucoma grading. It includes both 2D fundus images and 3D OCT images from several patients. Each sample is annotated with glaucoma grades alongside macular fovea coordinates and optic disc/cup segmentation masks for the fundus images. In the context of the challenge, the authors provided 100 accessible labeled samples and another 100 unlabeled cases as the benchmark.

\begin{table}[t]
\begin{center}
\caption{List of datasets supporting Glaucoma}
\label{tab:ds_glaucoma}
\setlength{\tabcolsep}{3pt}
\scalebox{0.8}{
\begin{tabular}{l|p{120pt}cccc}
\toprule
\textbf{Year} & \textbf{Dataset} &  \textbf{\# Images} & \textbf{Format} & \textbf{Resolution} \\
\toprule
2004 & ONHSD \cite{ONHSD2012} & 99 &-& 640$\times$480 \\ \midrule
2008 & Drions-DB \cite{Carmona:2008:ION:1383660.1383874} & 110 &-& 600$\times$400 \\ \midrule
2010 & ORIGA \cite{5626137}   & 650 &-& 3072$\times$2048 \\ \midrule
2011 & RIM-ONE \cite{5999143}   & 169 &-& 2144$\times$1424 \\ \midrule
2013 & ACHIKO-K \cite{6566371}   & 258 &-& - \\ \midrule
2015 & Drishti-GS \cite{Sivaswamy2015ACR}  & 101 &-& 2896$\times$1944 \\ \midrule
2018 & RIGA \cite{Almazroa2018}  & 750 &-& various \\ \midrule
2019 & LAG \cite{Li_2019_CVPR}   & 5,824 &-& - \\ \midrule
2020 & REFUGE \cite{ORLANDO2020101570}   & 1,200 &-& various \\ \midrule
\multirow{4}{*}{2023} & 
Rotterdam EyePACS AIROGS \cite{AIROGS2023}  & 113,893 & JPEG &-\\  \cmidrule{2-5}
& EyePACS-AIROGS-light \cite{10270429}  & 3,270 & JPEG & 256$\times$256\\ \cmidrule{2-5}
& EyePACS-AIROGS-light-v2 \cite{10.1145/3603765.3603779} &   4,770 & JPEG & 512$\times$512 \\ \cmidrule{2-5}
& Chákṣu \cite{Kumar2023} &  1,345 & JPEG/PNG & various \\
\cmidrule{2-5}
& GAMMA \cite{wu2023gamma} &  100 & 2D and 3D & various \\
\bottomrule
\end{tabular}
}
\label{tab1}
\end{center}
\end{table}

                                    
\subsection{\textbf{Age-related Macular Degeneration}}
\label{subsec:AMD}

\begin{table*}[t]
\begin{center}
\caption{Typical studies on the application of AI in Age-related Macular Degeneration}
\label{tab:amd_overview}
\setlength{\tabcolsep}{3pt}
\scalebox{0.8}{
\begin{tabular}{l|p{70pt} p{160pt} p{140pt} p{180pt}}
\toprule
\textbf{Year} & \textbf{Study} & \textbf{Method} & \textbf{Task} & \textbf{Performance} \\
\toprule
\multirow{2}{*}{2020}
& Jason et al. \cite{yim2020predicting}  & Integrating models: OCT images, tissue maps & AMD Classification &  Sen=0.8, Spec=0.55; Sen=0.34, Sen=0.90\\ \cmidrule{2-5}
& Yan et al. \cite{yan2020deep} &  Modified CNN &  AMD Progression & AUCROC=0.85 \\ \midrule
\multirow{2}{*}{2021} & 
 Hyungwoo et al. \cite{lee2021integrative} &  Unsupervised K-Means with PCA & Clustering: 5 drusen types & Significant differences in cluster parameters \\ \cmidrule{2-5}
& Pfau et al. \cite{pfau2021probabilistic}   & NGBoost & Predicting anti-VEGF injection frequency & MAE\_Lasso=2.76, MAE\_PCA=2.74, and MAE\_RF=2.6\\ \midrule

\multirow{4}{*}{2022} & 
Zahra et al. \cite{baharlouei2022detection} &  Wavelet scattering network and PCA Classifier & AMD Classification & Acc\_normal=0.974, Acc\_pathologies=0.825\\ \cmidrule{2-5}
& Han et al. \cite{han2022classifying}  & VGG-16, VGG-19, ResNet & AMD Classification &  Acc=0.874\\ \cmidrule{2-5}
& Jin et al. \cite{jin2022multimodal} & Unidirectional fusion network (UFNet) and the bidirectional fusion network(BFNet) & Anomaly Detection  & Acc=0.955, AUC=0.979\\ \cmidrule{2-5}
&  Lourdes et al. \cite{martinez2022explainable} & Artificial Hydrocarbon Networks & AMD Detection & Sen=0.989, Spec=0.989, Pre=0.987, F1=0.988\\
\cmidrule{2-5}
&  Zarauz-Moreno, Antonio, et al. \cite{zarauzhierarchical} & Hierarchical Transformer & AMD Classification & F1 = 72.11, Acc = 73.7, AUC = 94.42\\
\cmidrule{2-5}
&  Junghwan Lee et al. \cite{lee2022predicting} & CNN-LSTM, CNN-Transformer & AMD Progression Predict & AUC = 0.879 vs 0.868  for 2 years, AUC = 0.879 vs 0.862 for 5 years\\
\midrule
2023 & 
Mini et al. \cite{wang2023explainable} & VGG16 & AMD Detection & Acc=0.992\\ \bottomrule

\end{tabular}
}
\label{tab1}
\end{center}
\end{table*}

AI and DL present transformative opportunities for improving the diagnosis, monitoring, and treatment of AMD. By leveraging advanced imaging modalities like OCT, these technologies can detect subtle biomarkers, quantify pathological changes, and predict disease progression with high accuracy. However, for widespread clinical adoption, ensuring the interpretability and transparency of these AI models is as crucial as their accuracy. Clinicians need to trust and understand the decision-making process of these systems to integrate them into patient care effectively. This section delves into recent advancements in interpretable AI methods for AMD diagnosis, focusing on techniques that not only achieve robust diagnostic performance but also might provide insights from OCT imaging data. We summarize typical algorithms in Table \ref{tab:amd_overview}.

\vspace{0.1in}
\subsubsection{\textbf{Machine Learning Algorithms}}
The study by Baharlouei et al. \cite{baharlouei2022detection} introduces a low-complexity CAD system designed to classify retinal abnormalities in OCT images. Using a wavelet scattering network for feature extraction and principal component analysis (PCA) \cite{abdi2010principal} for classification, this system effectively detects conditions such as AMD, central serous retinopathy, DR, and macular holes. By automating multiclass classification, the proposed method can reduce the reliance on manual inspection by ophthalmologists, highlighting its practical value in clinical settings.

Building on the theme of leveraging AI for AMD management, Pfau et al. \cite{pfau2021probabilistic} propose an automated pipeline for predicting anti-vascular endothelial growth factor (anti-VEGF) treatment frequency in patients with neovascular AMD. Using volumetric spectral domain-OCT (SD-OCT) biomarkers and ML models such as natural gradient boosting (NGBoost) \cite{duan2020ngboost}, this system forecasts treatment needs over a 12-month period, providing a personalized and probabilistic approach to treatment planning in real-world settings.

Expanding the scope to nonexudative AMD subtypes, Lee et al. \cite{lee2021integrative} analyze structural parameters of Haller vessels and choriocapillaris (CC) using OCT and OCT angiography. This study quantifies vessel diameter, length, and intersections, as well as the total area and size of CC flow voids, revealing significant differences across AMD subtypes and pachydrusen. Notably, unsupervised machine learning \cite{celebi2016unsupervised,caron2018deep} identified four distinct clusters of eyes, highlighting variations in vascular characteristics among these groups and offering deeper insights into the pathophysiology of AMD.

\vspace{0.1in}
\subsubsection{\textbf{Convolutional Neural Networks}} 
\begin{figure}[b]
    \centering
    \includegraphics[width=0.48\textwidth]{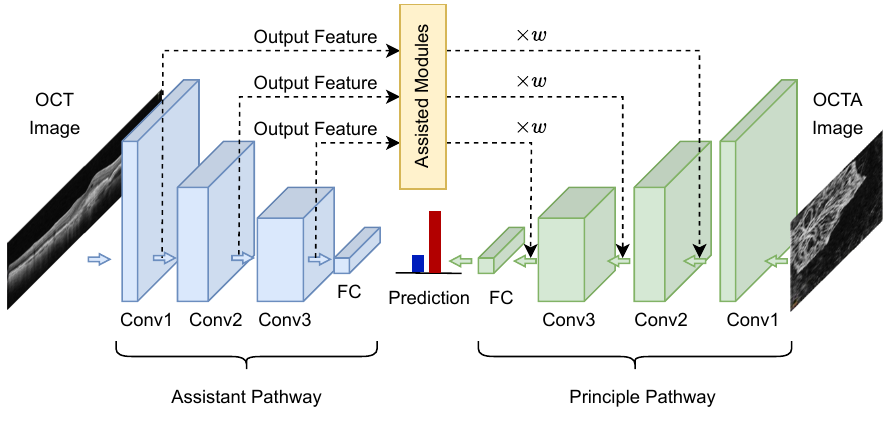}
    \caption{The main architecture of a unidirectional fusion network (UFNet) consists of feature pathways, assistant modules, and residual fusion. Attention feature maps generated from the Conv2-5 layer of the assistant pathway are provided for further fusion with the corresponding feature map generated by the principal pathway \cite{jin2022multimodal}.}
    \label{fig:jin2022multimodal}
\end{figure}

The research presented in \cite{yim2020predicting} 
explores the use of CNNs to predict the progression of exudative age-related macular degeneration (exAMD) in the second eye of patients already diagnosed in one eye. By leveraging automatic tissue segmentation, the research highlights the potential to detect early anatomical changes preceding conversion and identify high-risk subgroups, enabling proactive interventions.
Similarly, Jin et al. \cite{jin2022multimodal} investigate a model that integrates OCT and optical coherence tomography angiography (OCTA) data to assess choroidal neovascularization (CNV) in neovascular AMD. The study employs a novel feature-level fusion (FLF) method, combining outputs from unidirectional (UFNet) and bidirectional (BFNet) fusion networks. This dual-pathway approach, with OCT and OCTA images processed as primary and auxiliary inputs (illustrated in Figure \ref{fig:jin2022multimodal}), respectively, enhances the model’s capability to analyze multimodal data effectively.

Qi Yan et al. \cite{yan2020deep} present another approach by integrating genetic data with fundus images to predict the progression of late-stage AMD. Using a modified deep CNN and a dataset comprising 31,262 fundus images and 52 AMD-associated genetic variants from the Age-Related Eye Disease Study, the model achieved an impressive area-under-the-curve (AUC) of 0.85, outperforming image-only predictions. This demonstrates the added value of genetic data in refining disease risk assessments over long periods.

Focusing on AMD subtype differentiation, Han et al. \cite{han2022classifying} develop a CNN-based model to classify subtypes of neovascular AMD (nAMD) using spectral domain optical coherence tomography (SD-OCT) images. The model employs transfer learning \cite{pan2009survey,weiss2016survey,nguyen2022tatl} and data augmentation \cite{shorten2019survey} to enhance robustness and diagnostic precision. To address explainability in AMD diagnosis, Wang et al. \cite{wang2023explainable} suggest constructing a VGG16-based network with optimization and XAI techniques \cite{dwivedi2023explainable,angelov2021explainable} for interpretable AMD detection in medical IoT systems. Similarly, Martinez et al. \cite{martinez2022explainable} introduce the explainable artificial hydrocarbon networks (XAHN) explainer, transforming the artificial hydrocarbon networks model into a tree-based structure to provide global and local interpretability. The XAHN approach captures nonlinear feature interactions without relying on surrogate models, ensuring transparency and usability for clinicians.

Overall, these works highlight the transformative role of CNNs in advancing AI applications for AMD diagnosis and management. By leveraging CNN architectures tailored for tasks like multimodal imaging analysis, genetic data integration, and interpretable decision-making, these approaches enable more accurate, transparent, and patient-specific care strategies.

\vspace{0.1in}
\subsubsection{\textbf{Transformer-based Algorithms}}
Besides CNNs, transformer-based models have shown promising results in predicting AMD-related diseases. One notable approach is the use of hierarchical transformer models \cite{zarauzhierarchical} for classifying AMD from OCT images. These models utilize attention mechanisms to hierarchically process visual information, improving classification accuracy and reducing the number of trainable parameters. Another method involves combining ViTs with other DL architectures, such as ResNet, to predict the progression of AMD over time \cite{lee2022predicting}. This hybrid integration enhances the model’s sensitivity to subtle retinal changes indicative of disease advancement. These developments underscore the transformative potential of transformer-based models in advancing early detection, monitoring, and personalized management of AMD.

\vspace{0.1in}
\subsubsection{\textbf{Datasets}}
We summarize below the most commonly used datasets in ADM research.
\paragraph{AREDS-1} The AREDS-1 \cite{study1999age}  dataset, obtained from the National Eye Institute, initially comprised 188,006 images. These images underwent quality assessment using a neural network, resulting in 118,254 gradable images from 4,591 patients. Notably, 398 patients in the cohort developed advanced AMD during the AREDS study period.

\paragraph{ARIA} ARIA \cite{10.1371/journal.pone.0032435} (Automated Retinal Image Analyzer) incorporates algorithms for vessel detection and diameter measurement, and its associated database consists of 143 color fundus images (768$\times$576 pixels). The images are categorized into three classes: AMD subjects (n=23), healthy control-group subjects (n=61), and diabetic subjects (n=59).

\paragraph{KORA} The KORA dataset \cite{Brandl2016} is specifically curated to assess the prevalence of early and late-stage AMD features within a general adult population. It comprises fundus images collected from 2,840 participants aged 25 to 74 years, as part of the Cooperative Health Research in the Region of Augsburg project. This dataset provides a valuable resource for studying AMD across a diverse demographic.

\paragraph{OCTID} The open-access database \cite{GHOLAMI2020106532} with over 500 high-resolution images categorized into various pathological conditions such as Normal, Macular Hole (MH), AMD, Central Serous Retinopathy (CSR), and DR. The images were obtained using a raster scan protocol with a 2 mm scan length and a resolution of 512$\times$1024 pixels.

\paragraph{iChallenge-AMD} The iChallenge-AMD \cite{dt4f-rt59-20} dataset comprises 400 images, with 89 images from patients with AMD. Image sizes vary, with some at 2124$\times$2056 pixels and others at 1444$\times$1444 pixels. All images are manually labeled as AMD or non-AMD.

\begin{table}[H]
\begin{center}
\caption{List of datasets supporting AMD}
\label{tab:ds_amd}
\setlength{\tabcolsep}{3pt}
\scalebox{0.9}{
\begin{tabular}{l| p{80pt}cccc}
\toprule
\textbf{Year} & \textbf{Dataset} &  \textbf{\# Images} & \textbf{Format} & \textbf{Resolution} \\
\toprule
1999 & AREDS-1 \cite{study1999age} &188,006 &-  &-&  \\ \midrule
2012 & ARIA \cite{10.1371/journal.pone.0032435}  & 143 &-& 768$\times$576 \\ \midrule
2016 & KORA \cite{Brandl2016} &  2546 &-& 768 x 576 \\ \midrule
2018 & OCTID \cite{GHOLAMI2020106532}  & 500 &-& 512$\times$1024 \\ \midrule
2020 & iChallenge-AMD \cite{dt4f-rt59-20} & 400 &-& various \\ \bottomrule

\end{tabular}
}
\label{tab1}
\end{center}
\end{table}


\subsection{\textbf{Retinal Vessel Segmentation}}
\label{subsec:vessel}
The primary objective of retinal vessel segmentation is to facilitate the accurate and detailed analysis of retinal blood vessels, which is crucial for diagnosing and monitoring various ocular diseases. This process involves identifying and delineating the intricate network of blood vessels within the retina from retinal images (Figure \ref{fig:you2023cas}). In this context, we explore a range of state-of-the-art methodologies that leverage advanced (i) attention-based algorithms and (ii) encoder-decoder architectures. These approaches have shown significant promise in improving segmentation accuracy by focusing on the most relevant regions of the image and capturing complex patterns. The incorporation of attention mechanisms enables the model to selectively focus on crucial areas of the retinal image (such as blood vessels and the macula) without requiring uniform processing of the entire image. On the other hand, encoder-decoder frameworks are designed to transform the input into a compressed latent space and then reconstruct it into the output. In many cases, attention mechanisms are integrated into encoder-decoder architectures, allowing the model to prioritize specific regions of the data during both the encoding and decoding stages, thereby enhancing its ability to capture important features more effectively.

\vspace{0.1in}
\subsubsection{\textbf{Attention-based Algorithms}} 
In retinal vessel segmentation, attention mechanisms have proven to be essential for improving model accuracy by enabling selective focus on critical features while suppressing irrelevant ones. Dong et al. \cite{dong2022craunet} introduced the cascaded residual attention U-Net (CRAUNet), which integrates a multi-scale fusion channel attention (MFCA) module to improve vessel delineation by focusing on relevant features at different scales. Similarly, Li et al. \cite{li2022global} developed the GT-DLA-dsHFF model, which combines global transformer (GT) and dual local attention (DLA) mechanisms to capture both long-range dependencies and fine-grained local features. Liu et al. \cite{liu2023res2unet} further advanced this by proposing the DA-Res2UNet, using dual attention to adaptively highlight important regions while incorporating multi-scale feature extraction for more accurate segmentation. Ni et al. \cite{ni2023feature} also employed attention mechanisms in their FAF-Net model, focusing on the aggregation, reuse, and fusion of multi-scale features to minimize semantic information loss, further enhancing vessel segmentation accuracy.

In addition to these developments, other models have incorporated attention mechanisms in innovative ways to address specific challenges. Ouyang et al. \cite{ouyang2023lea} refined U-Net by adding a local feature enhancement module combined with an attention module, emphasizing relevant local features to improve segmentation precision. Transformer-based models, such as Shi et al.’s \cite{shi2023tcu} TCU-Net, utilize cross-fusion transformers and channel-wise cross-attention mechanisms to surpass traditional convolutional networks, offering more efficient segmentation by focusing on critical features and learning relationships between different image modalities. Wang et al. \cite{wang2023automatic} introduced a directed graph search-based method for vascular network segmentation, leveraging learnable attention to detect and prioritize feature points for more accurate results.

\begin{figure}[t]
    \centering
    \includegraphics[width=0.48\textwidth]{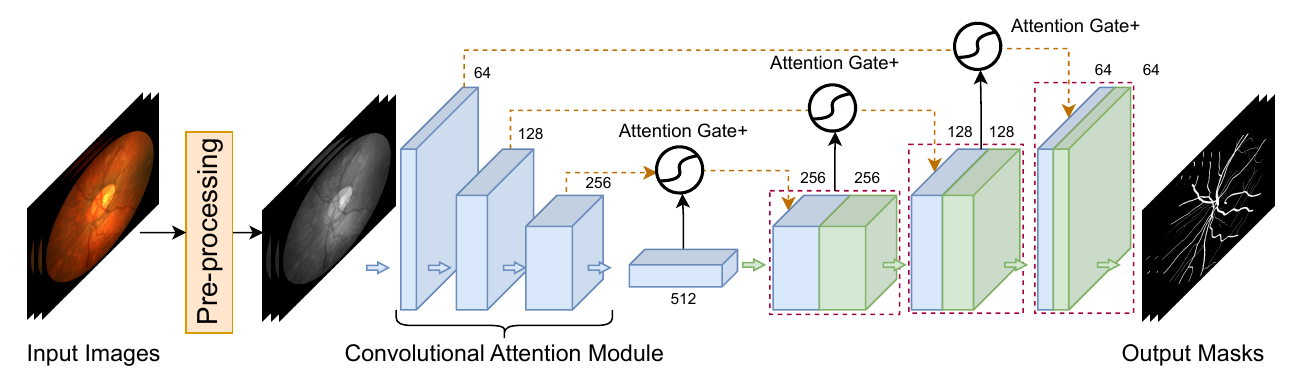}
    \caption{CAS-UNet architecture, it features Cross-Fusion Channel Attention and Structured Convolutional Attention blocks for channel enhancement, along with Additive Attention Gates in the skip-connection layer for spatial enhancement of retinal blood vessels \cite{you2023cas}.}
    \label{fig:you2023cas}
\end{figure}

Other notable advancements include You et al.’s \cite{you2023cas} CAS-UNet, which employs cross-fusion channel- and structured convolutional attention (Figure \ref{fig:you2023cas}), as well as an additive attention gate and soft pooling method, achieving high accuracy and sensitivity. Yuan et al. \cite{yuan2021multi} proposed AACA-MLA-D-UNet, incorporating an Adaptive Atrous Channel-aware mechanism to capture important features at multiple resolutions while reducing model complexity. This model also includes a multi-level module that improves generalization ability and performance across various retinal images. Together, these advancements demonstrate the growing importance of attention mechanisms in enhancing the robustness, accuracy, and generalization of retinal vessel segmentation models.

\begin{table*}[t]
\begin{center}
\caption{Typical studies on the application of AI in Vessel Segmentation.}
\label{tab:vessel_segment}
\setlength{\tabcolsep}{3pt}
\scalebox{0.8}{
\begin{tabular}{l|p{80pt} p{120pt} p{380pt} }
\toprule
\textbf{Year} & \textbf{Study} &  \textbf{Method} & \textbf{Performance} \\
\toprule
\multirow{3}{*}{2021} 
& Yukun et al. \cite{zhou2021learning}& Binary-to-multi-class Fusion Network& DRIVE: Sen=0.699, F1=0.700, ROC=0.841, HRF: Sen = 0.68, F1 = 0.72, ROC = 0.83\\ \cmidrule{2-4}
& Yuqian et al. \cite{zhou2021study} &  Concatenated UNet & DRIVE: Sen=0.838, Spec=0.983, DICE=0.832, Acc=0.971, AUC=0.989; CHASE\_DB1: Sen=0.869, Spec=0.984, DICE=0.827, Acc=0.977, AUC=0.992\\ \midrule

\multirow{4}{*}{2022} 
& Fangfang et al. \cite{dong2022craunet} &  CRAUNet & DRIVE\_AUC= 0.983, CHASE\_DB1\_AUC=0.987\\ \cmidrule{2-4}
& Li Yang et al. \cite{li2022global} &   Global Transformer (GT) and Dual Local Attention (DLA) network & DRIVE: Acc=0.970, Sen=0.836, Spec=0.983, AUC=0.986; STARE: Acc=0.976, Sen=0.848, Spec=0.986, AUC=0.991; CHASE\_DB1: Acc=0.976, Sen=0.844, Spec=0.986, AUC=0.989; HRF: Acc=0.969, Sen=0.817, Spec=0.983, AUC=0.985\\  \cmidrule{2-4}
& Xiang et al. \cite{zhong2022you}& Multi-layer multi-scale dilated convolution (MMDC-Net) network& STARE: Acc = 0.96, Sen = 0.85, Spe = 0.97, AUC = 0.97; CHASEDB1: Acc = 0.95, Sen = 0.84, Spe = 0.97, AUC = 0.95; DRIVE: Acc = 0.96, Sen = 0.81, Spe = 0.98, AUC = 0.96\\ \cmidrule{2-4}
& Rui et al. \cite{xu2022local}& Contrastive learning& DRIVE: ACC = 0.97, Sen = 0.84, Spe = 0.98, AUC = 0.99, Dice = 0.83; CHASE\_DB1: Acc = 0.98, Sen = 0.85, Spe = 0.99, AUC = 0.99, Dice = 0.82\\
\midrule

\multirow{8}{*}{2023} 
& Renyuan et al. \cite{liu2023res2unet}  & DA-Res2UNet & CHASE\_DB1: F1=0.819; DRIVE: F1=0.828; STARE: F1=0.839\\ \cmidrule{2-4}
& Jiajia et al. \cite{ni2023feature}   & Feature Aggregation and Fusion network (FAF-Net) & DRIVE: Acc = 96.08, Sen = 86.90, Spe = 97.37, AUC = 98.39; CHASE\_DB1: Acc = 97.53, Sen = 84.11, Spe = 98.43, AUC = 98.98; STARE: Acc = 97.10, Sen = 85.02, Spe = 98.36, AUC = 98.99 \\ \cmidrule{2-4}
&  Jihong et al. \cite{ouyang2023lea} & LEA U-Net &  DRIVE: Acc=0.9563, F1=0.823, TPR=0.7983, TNR=0.9793. The AUC of PRC is 0.9109, and the AUC of ROC is 0.9794\\ \cmidrule{2-4}
&  Zidi et al. \cite{shi2023tcu} &   TCU-Net & ROSE-1: Acc=0.945, AUC=0.862\\ \cmidrule{2-4}
&  Gengyuan et al. \cite{wang2023automatic} &  Directed graph search-based method & DRIVE: F1=0.863, Acc=0.914; IOSTAR: F1=0.764, Acc=0.854\\ \cmidrule{2-4}
&  Zeyu et al. \cite{you2023cas} & CAS-Unet&CHASE\_DB1: Acc=0.967, Sen=0.832; DRIVE: Acc=0.959, Sen=0.838\\ \cmidrule{2-4}
&  Jianyong et al. \cite{li2023gdf}& GDF-Net& CHASE\_DB1: Acc = 0.97, Sen = 0.79, Spe = 0.99, AUC = 0.99; STARE: Acc = 0.96, Sen = 0.76, Spe = 0.99, AUC = 0.99; DRIVE: Acc = 0.96, Sen = 0.83, Spe = 0.99, AUC = 0.99\\ \cmidrule{2-4}
& Yanhong et al. \cite{liu2023wave}&Wave-Net&DRIVE: Sen=0.816, Spec=0.976, Acc=0.956, F1=0.825\\ \cmidrule{2-4}
& Zijian et al. \cite{zijian2023affd} &   AFFD-Net & DRIVE: Sen=0.842; STARE: Sen=0.846; CHASE\_DB1: Sen=0.826\\ \bottomrule

\end{tabular}
}
\label{tab1}
\end{center}
\end{table*}
\vspace{0.1in}
\subsubsection{\textbf{Encoder-Decoder Algorithms}}
The publication \cite{li2023gdf} introduces GDF-Net, a multi-task symmetrical network designed for accurate retinal vessel segmentation. By employing two symmetrical segmentation networks, GDF-Net addresses information loss in thin vascular detection, capturing both global contextual and detailed features. The fusion network integrates these features, yielding improved segmentation accuracy as demonstrated in competitive experimental results.

Similarly, the publication \cite{liu2023wave} introduces Wave-Net, a lightweight model tailored for precise retinal vessel segmentation in fundus images. Overcoming challenges like semantic information loss and limited receptive field, Wave-Net incorporates a detail enhancement and denoising block (DED) and a multi-scale feature fusion block (MFF).
Moreover, \cite{xu2022local} introduces an enhanced U-Net model for retinal vessel segmentation, incorporating local-region and cross-dataset contrastive learning strategies. The local-region strategy focuses on separating features within local regions, while the cross-dataset strategy utilizes a memory bank scheme for global contextual information.

In a different approach, \cite{zhong2022you} introduces MMDC-Net, a multi-layer multi-scale dilated convolution network for retinal vessel segmentation. Addressing the lack of global information exploration and class imbalance, MMDC-Net employs an MMDC module and a multi-layer fusion module to capture blood vessel details effectively. The incorporation of a recall loss aids in resolving the class imbalance issue, demonstrating superior performance in terms of accuracy and sensitivity across various datasets. Additionally, \cite{zhou2021learning} presents a novel approach for multi-class segmentation of retinal blood vessels 
. Decomposing the segmentation task into binary classifications, including artery segmentation and vein segmentation, followed by a final multi-class prediction. By explicitly maintaining class-specific gradients and favoring discriminative features, this approach addresses intra-segment misclassifications in retinal imaging.


Study group learning (SGL) scheme is another direction proposed by Zhou et al. \cite{zhou2021study} to enhance the robustness of models trained on noisy labels for retinal vessel segmentation. This approach utilizes a concatenated U-Net architecture, which incorporates both enhancement and segmentation modules to process raw retinal images without requiring preprocessing. The SGL strategy partitions the training set into multiple subsets, allowing individual models to be trained independently. By aggregating knowledge from these distinct subsets, SGL helps reduce overfitting to noisy labels. Additionally, the Vessel Label Erasing technique simulates incomplete annotations, further enhancing the model’s ability to segment smaller objects accurately.

Building on similar objectives of improving segmentation accuracy and generalization, the AFFD-Net proposed by Zijian et al. \cite{zijian2023affd} introduces a dual-decoder network to address challenges such as low sensitivity and poor generalization in retinal vessel segmentation. The model incorporates modifications like reduced convolution filters and additional modules for multi-scale feature extraction, which contribute to improved sensitivity and segmentation performance. When evaluated on public databases, AFFD-Net outperforms classical networks, demonstrating superior generalization and segmentation accuracy with fewer parameters. 

Overall, the aforementioned advancements, summarized in Table \ref{tab:vessel_segment}, highlight continuous efforts to enhance the efficiency and effectiveness of retinal vessel segmentation models, particularly through the use of encoder-decoder and attention-based methods. These approaches are especially effective in addressing challenges such as noisy data and complex image features, significantly improving model performance in demanding segmentation tasks.
\vspace{0.1in}
\subsubsection{\textbf{Datasets}}
\paragraph{STARE} The STARE \cite{845178} database, generated by scanning and digitizing 20 retinal image photographs, has lower image quality compared to other public databases. Captured by a narrow field of view (35 degrees) camera, the images in the STARE database have a resolution of 700$\times$605 pixels.

\paragraph{DRIVE} The DRIVE \cite{staal2004ridge} database comprises 40 retinal images, with 33 depicting healthy conditions and 7 exhibiting specific pathologies. Captured with a fundus camera featuring a 45-degree field of view, the images in this database have a resolution of 565$\times$584 pixels.

\paragraph{CHASE\_DB1} Kingston University, London, in collaboration with St. George’s, University of London, has released a public retinal vessel reference dataset, CHASE\_DB1 \cite{6224174}, comprising 28 retinal images from multi-ethnic children in the Child Heart and Health Study in England (CHASE). The images in this database have a resolution of 1280$\times$960 pixels.

\paragraph{HRF} The HRF \cite{wang2013robust} dataset, designed for retinal vessel segmentation, consists of 45 images arranged into 15 subsets. Each subset includes a healthy fundus image, an image of a patient with DR, and a glaucoma image, with image sizes set at 3,304$\times$2,336 pixels.

\paragraph{IOSTAR} The IOSTAR \cite{IOSTAR2015} vessel segmentation dataset consists of 30 retinal images, each with a resolution of 1024$\times$1024 pixels. Expert annotations cover vessel segmentation, optic disc, and artery/vein ratio.

\paragraph{RC-SLO} The RC-SLO \cite{rcslo2015} dataset comprises 40 image patches, each with a resolution of 360$\times$320 pixels. It includes vessel annotations by retinal image analysis experts and covers various challenging scenarios like high curvature changes, central vessel reflexes, micro-vessels, and crossings/bifurcations.

\paragraph{ROSE-1} The ROSE-1 \cite{ma2021rose} encompasses 117 OCTA images from 39 subjects, with 26 having diseases and the remainder serving as healthy controls. All OCTA scans were acquired using the RTVue XR Avanti SD-OCT system by Optovue (USA) with AngioVue software, boasting an image resolution of 304$\times$304 pixels.

\begin{table}[!hbt]
\begin{center}
\caption{List of datasets supporting Retinal Vessel Segmentation.}
\vspace{0.05in}
\label{tab:ds_vessel}
\setlength{\tabcolsep}{3pt}
\scalebox{0.95}{
\begin{tabular}{l|p{80pt}ccc}
\toprule
\textbf{Year} & \textbf{Dataset} &  \textbf{\# Images} & \textbf{Format} & \textbf{Resolution} \\
\toprule
2000 & STARE \cite{845178} &  20 & JPEG & 700$\times$605 \\ \midrule
2004 & DRIVE \cite{staal2004ridge} &  40 & JPEG & 565$\times$584 \\ \midrule
2009 & CHASE\_DB1 \cite{6224174}  & 28 & TIFF & 1280$\times$960 \\  \midrule
2013 & HRF \cite{wang2013robust}& 45 & - & 3304$\times$2336 \\ \midrule
2015 & IOSTAR \cite{IOSTAR2015} &  30 & JPEG & 1024 $\times$1024 \\ \midrule 
2015 & RC-SLO \cite{rcslo2015} &  40 & JPEG & 360$\times$320 \\ \midrule 
2021 & ROSE-1 \cite{ma2021rose}& 117 & - & 304$\times$304 \\ 
\bottomrule
\end{tabular}
}
\label{tab1}
\end{center}
\end{table}

\section{\textbf{Future Perspectives}}
\label{sec:future}
\begin{figure*}[t]
    \centering
    \includegraphics[width=0.8\textwidth]{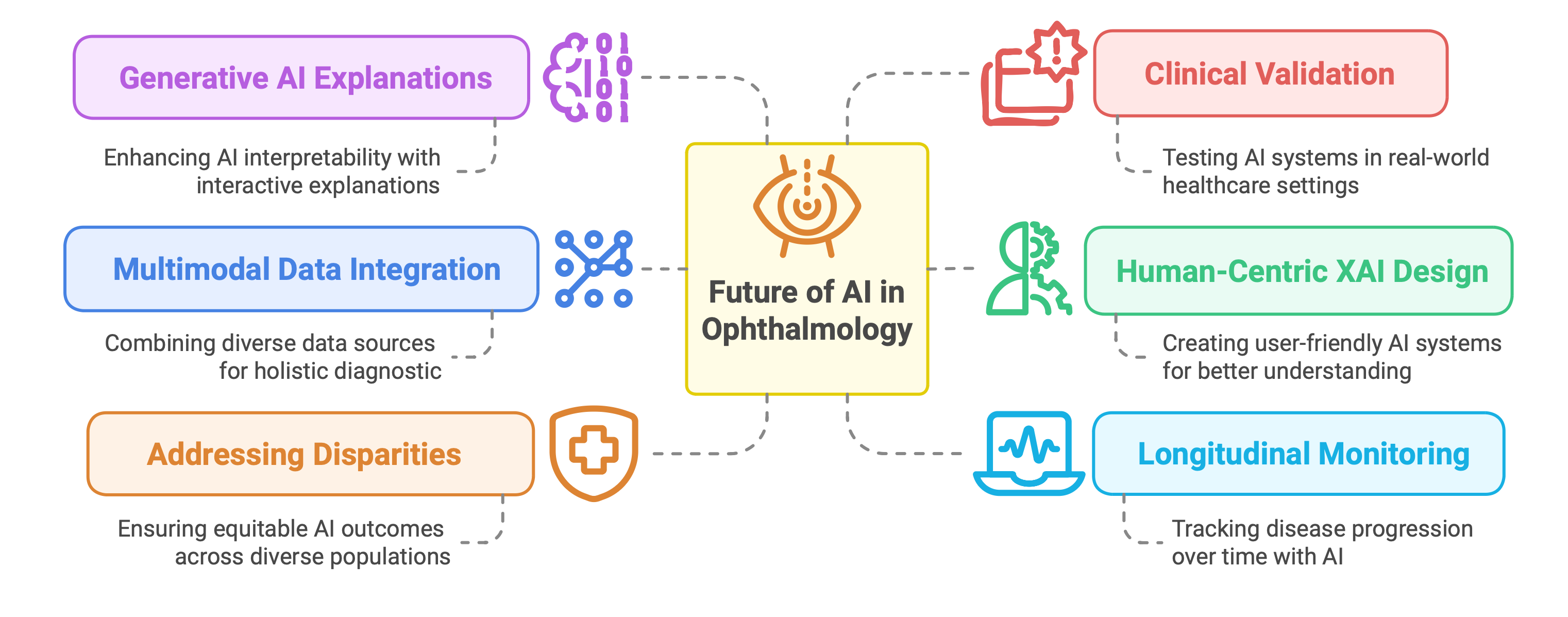}
    \caption{Future perspectives in AI for Ophthalmology encompass several key areas, including addressing disparities, integrating multi-modal data, enhancing generative AI explanations, validating models in real-world clinical settings, focusing on human-centric explainable AI (XAI), and enabling longitudinal monitoring.}
    \label{fig:future_directions}
\end{figure*}

The field of medical imaging, particularly in diagnosing eye diseases such as DR, macular edema, AMD, glaucoma, and retinal vessel segmentation, has experienced remarkable progress, with advancements driven by XAI and human-in-the-loop methodologies. These approaches have significantly enhanced the transparency and trustworthiness of AI models, making them more acceptable in clinical settings. At the same time, DL techniques, including both CNNs and transformer-based architectures, have demonstrated exceptional capabilities in extracting intricate patterns and features from complex medical images. Despite these advancements, there are still numerous promising opportunities for future research to enhance the interpretability, accuracy, and practical applicability of AI-driven solutions in ophthalmology. This section explores potential future directions for advancing the development of more reliable, efficient, and clinically adaptable AI-based models in ophthalmic applications. A visual representation of these directions is provided in Figure \ref{fig:future_directions}.

\subsection{\textbf{Integration of Multimodal Data}}
One promising direction in advancing AI-driven diagnostics for ophthalmology is the integration of diverse data sources beyond conventional retinal imaging. By incorporating genetic markers, patient demographics, lifestyle factors, and clinical histories, a more holistic view of the pathogenesis and progression of eye diseases can be achieved. For instance, the role of genetic markers has been emphasized in studies such as \cite{abd2019developing}, which highlight the potential of leveraging genomics for identifying disease susceptibility. Similarly, patient demographic information, such as age, ethnicity, and socioeconomic status, has proven valuable in risk stratification, as demonstrated by \cite{alanazi2022identification}. Lifestyle factors, including diet and smoking habits, play a significant role in diseases like DR and macular degeneration, as explored in \cite{ajana2021predicting}. Additionally, the integration of longitudinal clinical histories can provide temporal insights into disease progression, as evidenced by \cite{hong2019medical}.

This multimodal approach, which combines diverse data modalities \cite{nguyen2023joint}, enables AI systems to capture the intricate interplay of genetic predispositions, environmental exposures, and disease trajectories, leading to more personalized and precise diagnostic assessments \cite{singh2022novel}. The use of advanced multimodal fusion techniques and machine learning algorithms on this aspect is crucial in extracting meaningful insights from such heterogeneous datasets. Techniques like attention-based fusion \cite{sharma2021medfusenet,zhou2022multi,wang2024advances}, graph neural networks \cite{choi2017gram,li2022graph,zhu2021survey}, and large-scale pre-trained medical models \cite{wang2022medclip,mh2024lvm,ma2024segment,zhao2024foundation} can effectively model relationships between disparate data types, enhancing diagnostic accuracy and robustness. Furthermore, these approaches improve interpretability, enabling clinicians to understand how different data sources contribute to diagnostic decisions. 

\subsection{\textbf{Human-Centric Design of XAI Systems}}
Future research should prioritize the design and implementation of XAI systems that are grounded in user-centric principles, as emphasized in \cite{calisto2021introduction,jin2022artificial}. Such systems are particularly important in bridging the gap between advanced AI models and their practical utility in clinical environments, particularly in diagnosing and managing eye diseases such as DR, AMD, and glaucoma. Grounded in user-centric principles, these systems have to deliver outputs that are comprehensible and actionable for healthcare professionals. For example, interactive tools like saliency maps or heatmaps \cite{selvaraju2017grad,ayhan2022clinical} over retinal images can help ophthalmologists pinpoint areas critical to AI predictions, fostering trust and improving diagnostic accuracy. Seamless integration into existing workflows through intuitive dashboards and real-time visualization of AI-based support systems \cite{bjerager2024real} can further enhance usability and adoption.

Additionally, integrating clinician feedback mechanisms into XAI systems \cite{holzinger2017we,tusfiqur2022drg,prentzas2023explainable} can empower ophthalmologists to refine model outputs and validate predictions, leading to better alignment with clinical needs. These mechanisms allow AI systems to dynamically adapt to specific diagnostic contexts, improving accuracy and reliability. Furthermore, XAI systems should clearly communicate uncertainties in their predictions \cite{yang2021uncertainty,zou2023review,nguyen2023out,laves2020well}, enabling clinicians to make informed decisions, especially in cases with ambiguous findings. By prioritizing transparency, interactivity, and clinician involvement, XAI can become a transformative tool in ophthalmology, facilitating precise, efficient, and trustworthy patient care.

\subsection{\textbf{Longitudinal Monitoring and Prognostic Modeling}}
Expanding beyond static diagnostic tasks, the future direction in ophthalmology should focus on longitudinal monitoring and prognostic modeling of eye disease progression. By analyzing temporal changes in retinal morphology, vascular patterns, and other biomarkers, AI systems using transformer \cite{holste2024harnessing} or graph neural networks \cite{emre20243dtinc,el2024eye} can provide early warnings of disease exacerbation and guide personalized treatment strategies. For example, tracking retinal thickness or vascular anomalies over time could enable early detection of DR progression or macular edema recurrence.

Prognostic models with interpretability can empower clinicians to make informed decisions about patient management and treatment planning, such as predicting responses to therapies like anti-VEGF in AMD degeneration. Similar approaches have proven effective in other fields, like Alzheimer’s disease, where machine learning forecasts cognitive decline from longitudinal data \cite{zhu2021long}. These advancements hold promise for improving both preventive and therapeutic outcomes in ophthalmic care.

\subsection{\textbf{Interactive Explanations with Generative AI}}
A compelling avenue for future research is the integration of Generative AI, particularly large language models (LLMs), into ophthalmology applications to enhance the interpretability and usability of AI-driven diagnostic systems. Models such as GPT \cite{openaiGPT4}, Gemini \cite{googleGeminiChat}, LLaMA \cite{metaLlamaMeta}, Claude \cite{claudeClaude}, Mixtral \cite{mistralMixtralExperts}, and Falcon \cite{tiiFalcon}, along with domain-specific adaptations like LLaVA-Med \cite{li2024llava}, BioMed-GPT \cite{zhang2024generalist}, and LoGra-Med \cite{nguyen2024logra}, hold significant potential for creating interactive explanations of AI predictions. These models can generate human-readable explanations as demonstrated in \cite{wang2023r2gengpt}, tailored to individual users' understanding and preferences, enhancing transparency and trust in AI-driven diagnostic systems. Empowering end-users, including healthcare professionals and patients, to interact dynamically with LLM-generated explanations fosters collaborative decision-making and enhances understanding of complex diagnostic outcomes. This interactive approach also allows clinicians to delve deeper into the AI’s reasoning, improving confidence in the system while enabling patients to comprehend their diagnoses and treatment options better \cite{wu2024guiding,chen2024huatuogpt}.

To ensure safety and mitigate potential distress for patients, integrating the principles of constitutional AI (CAI) \cite{bai2022constitutional} offers a valuable framework for designing AI-driven diagnostic systems. CAI emphasizes transparency, non-evasiveness, and harmlessness in AI interactions, guiding systems to produce responses that are both informative and empathetic. For instance, AI models can present diagnostic findings in a manner that is clear yet reassuring, reducing patient anxiety and fostering trust in the AI-assisted diagnostic process. This thoughtful integration of ethical principles ensures that AI systems not only enhance clinical decision-making but also support a patient-centered approach in healthcare.

\subsection{\textbf{Addressing Disparities and Bias}}
Mitigating disparities and biases in AI-driven diagnostic systems should be a central focus of future research to ensure equitable healthcare for all patients. A critical aspect of this is the development of fairness-aware algorithms \cite{tian2025fairdomain,xu2022algorithmic,zong2022medfair} that actively identify and reduce demographic biases, ensuring that AI models deliver consistent, unbiased diagnostic outcomes across diverse patient populations. These algorithms can address biases related to age, gender, ethnicity, and socioeconomic status, helping to create more equitable AI tools. In parallel, incorporating diverse and representative training datasets is essential to improve the generalizability of AI models. By including data from a wide array of geographic regions, ethnic groups, and clinical settings \cite{drabiak2023ai,chen2021ethical}, AI systems can better reflect the variety of patient experiences and disease manifestations. This not only reduces algorithmic biases but also enhances the robustness and accuracy of AI models across different populations. Additionally, standardizing data collection protocols and encouraging the sharing of underrepresented datasets can further strengthen the inclusiveness of these systems. Collectively, these efforts will contribute to more fair and reliable AI-driven diagnostic tools that promote trust and equality in healthcare

\subsection{\textbf{Clinical Validation and Real-World Deployment}}

Lastly, prospective research should prioritize rigorous clinical validation and real-world deployment of XAI-driven diagnostic systems to ensure their effectiveness in diverse healthcare settings \cite{skevas2024implementing,vaughan2024review,chen2024towards}. Collaborative studies involving multi-center clinical trials and real-world implementation in diverse healthcare settings are essential, as they can provide valuable insights into the scalability, effectiveness, and clinical utility of AI-driven solutions. Furthermore, continuous monitoring of system performance and user feedback in these real-world settings will enable iterative improvements and optimization of XAI algorithms, 
ultimately ensuring that AI-driven tools are not only scientifically robust but also practical and beneficial for clinical use.



\section{\textbf{Conclusion}}
\label{sec:conclusion}
In this survey, we introduce a comprehensive overview of the current state-of-the-art AI methods, with a particular focus on DL approaches such as CNNs and transformer architectures. We explore their applications in major ophthalmic conditions, including DR, glaucoma-related diseases, AMD, and retinal vessel segmentation. Building on this overview, we highlight key areas for future research, emphasizing the importance of interdisciplinary collaboration, user-centric design, longitudinal monitoring, interactive explanations, fairness, and real-world validation. We expect that this survey will provide valuable insights into the current landscape of AI in ophthalmology and inspire further research to address ongoing challenges, ultimately unlocking the full potential of AI for improving diagnosis and patient care.

\renewcommand{\bibfont}{\normalfont\footnotesize}
\printbibliography[resetnumbers=true]

\end{document}